\documentclass[12pt]{article}
\usepackage[margin=1in]{geometry}
\usepackage{graphicx}
\usepackage{bm}
\def\Vec#1{{\bm #1}}
\def\lesim{\ \hbox to 0 pt{\raise .6ex\hbox{$<$}\hss} \lower.5ex\hbox{$\sim$}\ }
\def\gesim{\ \hbox to 0 pt{\raise .6ex\hbox{$>$}\hss} \lower.5ex\hbox{$\sim$}\ }

\title{Particle in cell calculation of plasma force on a small grain
  in a non-uniform collisional sheath}

\author{I H Hutchinson}
\date{Plasma Science and Fusion Center and\\ Department of Nuclear
  Science and Engineering,\\ Massachusetts Institute of Technology,\\
  Cambridge, MA, USA}
\begin{document}
\maketitle

\begin{abstract}
  The plasma force on grains of specified charge and height in a
  collisional DC plasma sheath are calculated using the multidimensional
  particle in cell code COPTIC. The background ion velocity distribution
  functions for the unperturbed sheath vary substantially with
  collisionality. The grain force is found to agree quite well with a
  combination of background electric field force plus ion drag force. 
  However, the drag force must take account of the non-Maxwellian (and
  spatially varying) ion distribution function, and the collisional
  drag enhancement. It is shown how to translate the dimensionless
  results into practical equilibrium including other forces such
  as gravity.
\end{abstract}

\section{Introduction}

Many dusty plasma phenomena are associated with the suspension of dust
grains near the edge of a sheath formed between the plasma and a wall,
e.g.\cite{Melzer1994,hebner03,Samarian2005,Fortov2005}. The grains therefore
reside in an inherently non-uniform plasma environment, and the scale
length of non-uniformity is generally comparable to that of the
shielded grain potential. The ``plasma'' forces on the grain consist
of the ion drag arising from the ion flow into the wall-sheath, and
the electric field force from the sheath potential gradient. In
addition, gravity and other forces such as neutral drag or
thermophoretic force may need to be considered, but they are not the
object of the present calculations. The ion drag force is the quantity
least well established. The present work is devoted to calculating by
direct particle in cell (PIC) simulation the total plasma force
including the ion drag on a grain in a self-consistent sheath.  The
results provide quantitative theoretical total plasma force values and
establish the extent to which the inherent non-uniformity affects the
result, by comparing the force with that derived from formulas for ion
drag force in \emph{uniform} plasmas.

The physics of a collisional sheath is well
established
\cite{Riemann1991,Riemann2003b}
and the non-linear kinetic equations have been solved
\cite{Riemann1981,Vasenkov2002a,Jelic2007}
in one dimension. To calculate the drag force on a grain by
simulation, however, requires a multidimensional calculation, because
the drag arises from the nearby perturbation of the plasma. A grain
finite in the transverse dimensions breaks the one-dimensional
symmetry. At least two-dimensions in space are required and three in
velocity. In fact the present work is performed using a code that is
three-dimensional in space, COPTIC\cite{Hutchinson2011a}. The extra
computational effort of an additional dimension, while significant, is
partly recouped by the averaging inherent in obtaining the
longitudinal force. A fully three-dimensional code is required to
explore the interaction of multiple grains, for which COPTIC is fully
equipped\cite{Hutchinson2011b}, but that endeavor will not be
undertaken here.

The organization of the paper is that section \ref{colnsheath}
describes the methods of calculation in the context of the bare,
one-dimensional sheath, which provides the background plasma in which
the grain is to be placed. Three different levels of collisionality are
explored. Section \ref{grainforce} explains how the force on the grain
is expressed and calculated, and gives the numerical results for a
range of grain charge and height in the sheath. Then section
\ref{comparisonuniform} compares these results with what would be
predicted by combining the electric field force and the theoretical
ion drag force for a \emph{uniform} plasma having parameters equal to
their local values in the non-uniform sheath. The discussion section
illustrates the application to a characteristic experimental situation.

\section{Collisional Sheath}
\label{colnsheath}

The first requirement of an accurate theoretical calculation of grain
force in a sheath is an accurate representation of the
sheath. Strictly speaking, the interest is in the regions spanning the
sheath and the presheath, where the plasma makes a transition from the
quasineutral presheath (or some would say simply the ``plasma'') into
the region of non-negligible charge density that is the sheath. It is well
known that a slab-geometry non-uniform presheath cannot exist without
particle, momentum, or energy sources in the presheath%
\cite{Riemann1981,hutchinson1987,Riemann1991}
. They arise from collisions.

In this paper we exclude ionization, which may be important in many
physical situations.  We also ignore the possibility of strong neutral
density gradients or flow arising from recycling and atomic processes
close to the wall.  We allow only for charge-exchange ion collisions
with a uniform, stationary, background neutral gas.  They are quite
well represented by a Bhatnagar, Gross and Krook (BGK) type of
collision term. The collisions are therefore substantially
idealized. A collision consists of replacing the colliding ion with a
new one having velocity drawn from the neutral atom distribution
function. Although the charge exchange cross-section for commonly used
gases is not exactly inversely proportional to inter-particle speed,
it is a reasonable approximation to take the collision frequency to be
independent of speed. This approximation produces substantial
simplifications, such as making the background solution of the
Boltzmann equation separable, but does not represent precisely the
collisional atomic physics.  The neutral distribution is here taken to
be a stationary Maxwellian with temperature $T_n$.
\begin{figure}[htp]
  \centering
\includegraphics[width=10cm]{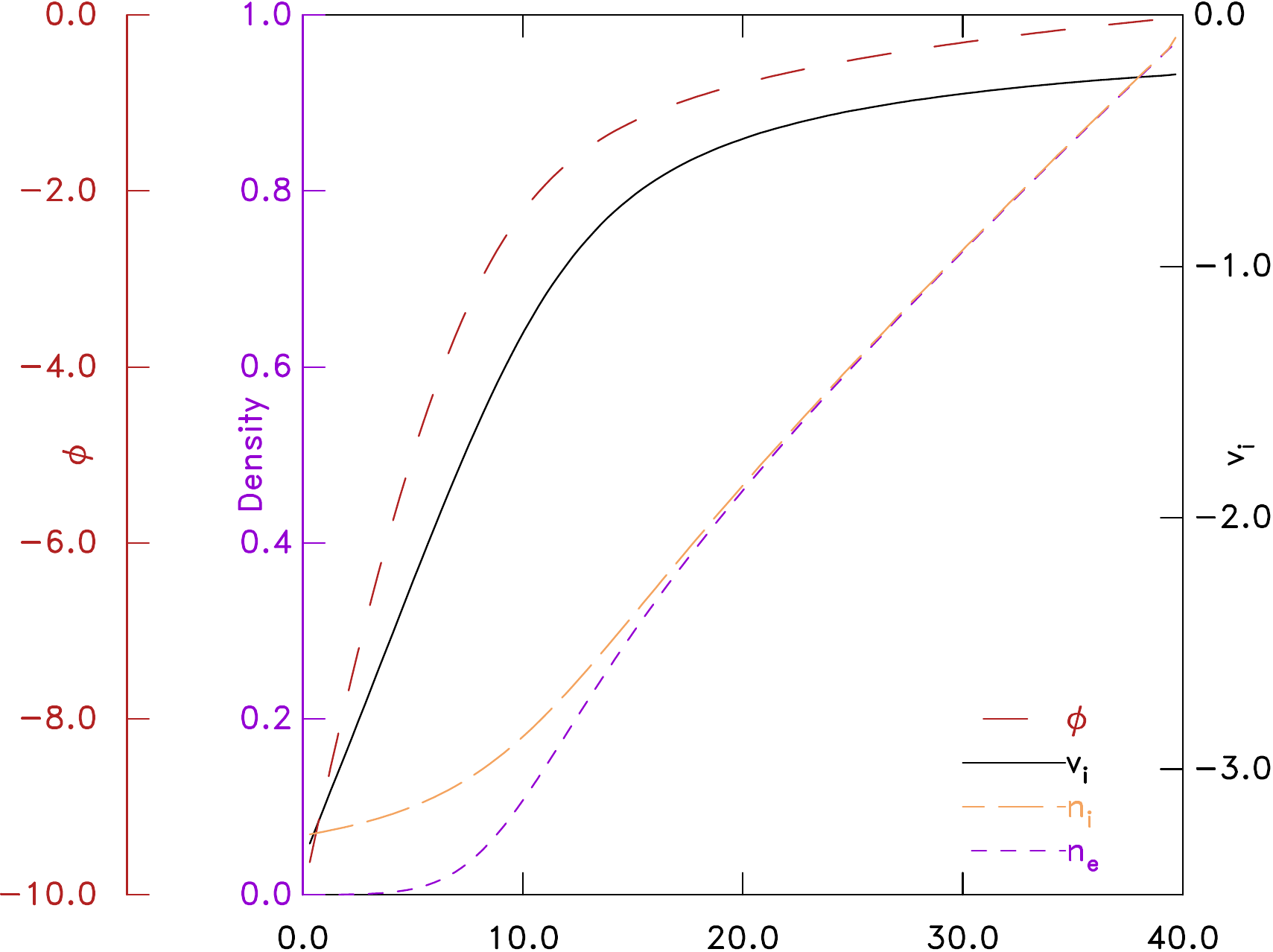}
  \caption{Collsional sheath spatial structure for collision time
    $\tau_c= 10$. Spatial distances are in units of the electron
    Debye length at the reference position, $z=40$. The wall is at
    $z=0$. All quantities are in normalized units, with $v_i$ denoting
    the fluid ion flow speed ($v_f$).}
  \label{sheathprofilesct10}
\end{figure}

\paragraph{Moderate collisionality.} The form of the presheath/sheath structure is illustrated in Fig.\
\ref{sheathprofilesct10}. Four quantities are plotted: the electron
and ion densities $n_e$ and $n_i$, the electric potential $\phi$ and
the ion fluid (average) velocity, $v_f$, all as a function of distance
$z$ from the wall. All quantities are expressed in normalized units
defined with respect to the reference position $z=40$. The units of
$z$ are electron Debye lengths $\lambda_{De}= \sqrt{\epsilon_0T_e/n_{e40}
  e^2}$, and of potential are $T_e/e$, where $T_e$ is the assumed
uniform electron temperature. The units of velocity are the cold-ion
sound speed $c_s=\sqrt{T_e/m_i}$, and the densities are normalized to
the reference density. Collisions occur at a rate defined by a
collision time $\tau_c$ expressed in units of
$\lambda_{De}/c_s$. Therefore, the value $\tau_c=10$ makes the mean
free path at speed $c_s$ equal to $10$ (normalized space units), a
quarter of the space range plotted. We will consider only
$T_n=0.02T_e$ in this paper.

The general character is that, starting from a distant position in the
presheath at the right, the density decreases towards the wall. The
potential and the electron density are presumed to be governed by a
Boltzmann relationship:
\begin{equation}
  \label{Boltzmann}
  n_e(\phi)= n_{e40}\exp(e\phi/T_e),
\end{equation}
or in normalized units $n_e=\exp(\phi)$. This approximation not
significantly affected by electron loss (cutting off the distribution)
far from the wall, because virtually all electrons there are
reflected. Although distribution cut-off compromises its accuracy
close to the wall, the inaccuracy does not matter because the
electron density is negligible there anyway.  As the potential drops, the
ions accelerate from subsonic speed. As they approach the sound speed
$v_f=-1$, quasi-neutrality breaks down and the electron and ion
densities begin to differ, and non-negligible positive charge density is
present. This is the entrance into the sheath. In the sheath the
potential falls more steeply, soon reducing the electron density to a
negligible level, while the ions continue to accelerate to supersonic
speeds through the potential gradient.

The wall in these calculations is considered to be held at a constant
(negative) potential $\phi=-10$, and the relevant solution is
steady. This formally represents a DC sheath. RF sheaths are
frequently of practical significance and have been studied in the
context of dust levitation both analytically\cite{Nitter1996} and
computationally\cite{ikkurthi10}. They act approximately like a DC
sheath with the ions feeling the time-averaged potential, but with the
average electron current to the wall adjustable by the amplitude of
the RF component of the potential. Consequently RF sheaths can
experience controllable, and sometimes quite negative, wall
potentials. Although RF modulation of atomic excitation is observed in
experiments, this is attributed mostly to energetic electrons; and the
grain charge is barely modulated\cite{Melzer2011}. The simulation
ignores all modulation and effectively considers the wall potential to
be controllable. If that potential is made more negative, all that
happens is that the solution shape remains unchanged but it moves
horizontally toward the right. Therefore a single simulation is
sufficient to represent virtually any wall potential provided we do
not wish to explore positions closer to the wall.

Figure \ref{sheathprofilesct10} is the result of an actual Particle in
Cell calculation using the code COPTIC, which is now explained.  Ions
are represented by individual particles (up to 25 million total)
governed by Newton's law of motion (in time units normalized to
$\lambda_{De}/c_s$)
\begin{equation}
  \label{Newton}
  {d \Vec{v}\over dt} = -\nabla \phi.
\end{equation}
The particles will be referred to as ions. Strictly speaking they are
superparticles, each representing a large number of ions, but most
intuitions based on regarding them as ions are physically correct.
At each time-step their velocities and positions are updated (using a
leapfrog scheme). Collisions possibly interrupt an ion's time-step
as determined statistically. If so, the remainder of the step with the
random new velocity is treated as a shortened time-step (possibly
itself to be interrupted).  The resulting ion density, when every
ion's step is completed, is deposited on a (non-uniform, typically
$32\times32\times128$) cartesian mesh of cells, and the new potential
is found by solving the finite difference form of
\begin{equation}
  \label{potentialeq}
  \nabla^2 \phi = n_e -n_i.
\end{equation}
The system is time-stepped until it converges (up to 6000 steps of
length $dt=0.1$, though with shorter timestep during measurement
periods and close to a grain). 
The boundary conditions used are as follows. Potential is fixed at the
wall $z=0$, periodic on the transverse ($x$ and $y$) boundaries, and its
gradient is fixed $d\phi/dz=\tau_c/|v_{f40}|$ at the boundary $z=40$. The
parameter $v_{f40}$ is the flow speed at $z=40$. It can be considered
an eigenvalue. There is a self-consistent solution only for one value
of $v_{f40}$. That value is found by iteration. For $\tau_c=10$, it is
$v_{f40}=-0.23$. [Note that a (non-ionizing) collisional sheath does
not have a constant-density asymptote far from the wall. Instead it
has velocity inversely proportional to a linearly rising density.]

The ions are treated as follows at the boundaries. The $x$ and
$y$ boundaries are periodic: particles that leave are reintroduced at
the opposite face. The wall just absorbs ions. None are injected
there. The ions crossing the plasma boundary $z=40$ are removed, but
other ions are injected at a constant rate and with a distribution
function that represents the appropriate collisional drift
distribution for the flow speed $v_{f40}$ (and density
$n_i=1$). That distribution\cite{Patacchini2008,Lampe2012} can be
written
\begin{equation}
  \label{driftdist}
  f_i({\bf v}) = {n_i\over \pi v_{tn}^2}{1\over 2v_f} \exp\left(-{v^2\over v_{tn}^2}\right) {\rm erfcx}
\left({v_{tn}\over 2 v_f} - {v_z\over v_{tn}}\right) ,
\end{equation}
where $v_{tn}=\sqrt{2T_n/m_i}$ and  ${\rm
  erfcx}(x) \equiv \exp(x^2){\rm erfc}(x)$.
\begin{figure}[htp]
\noindent
\vbox{\noindent
\hsize=0.48\hsize\includegraphics[width=\hsize]{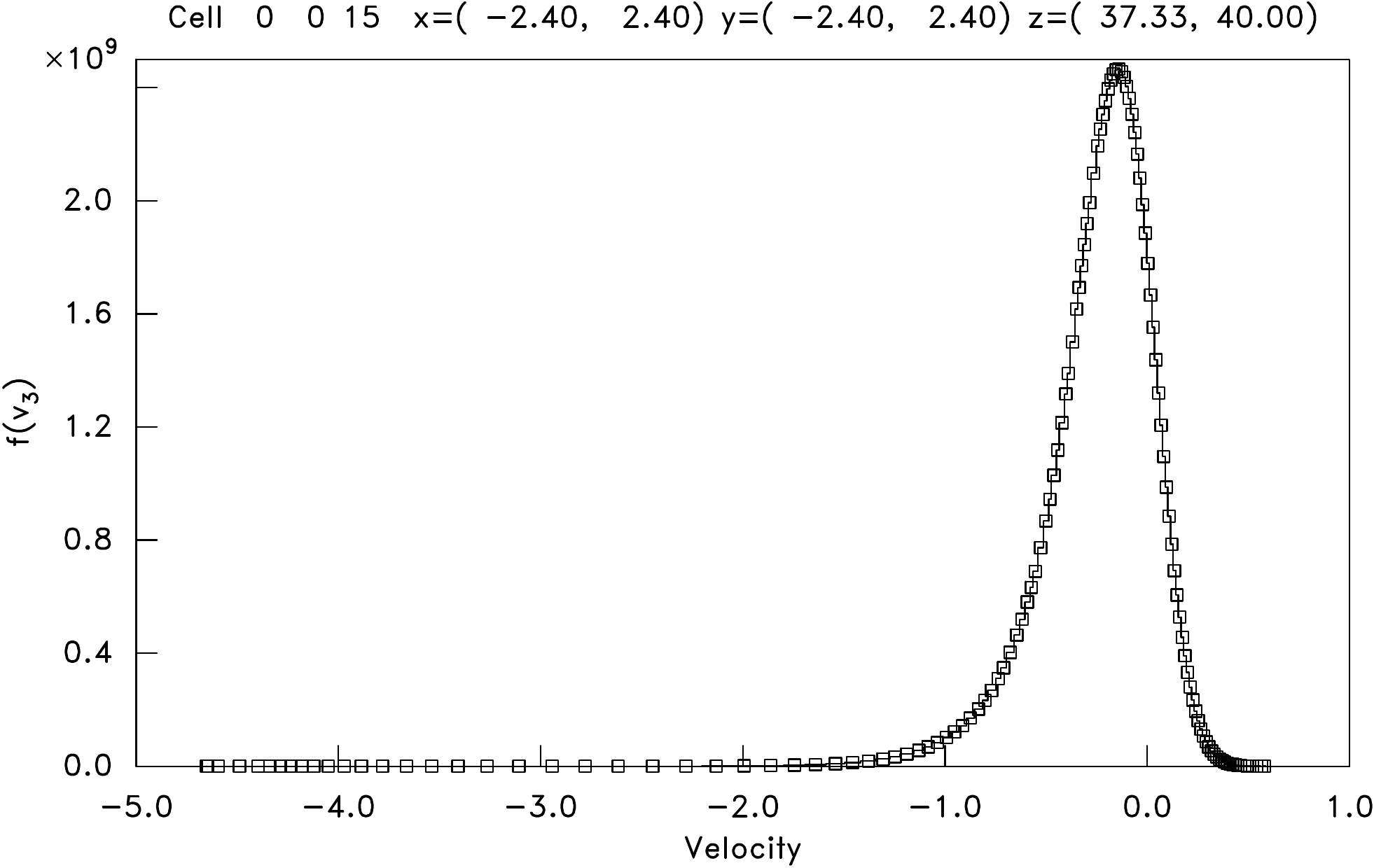}\par\vskip-10pt (a)}
\vbox{\noindent
\hsize=0.48\hsize\includegraphics[width=\hsize]{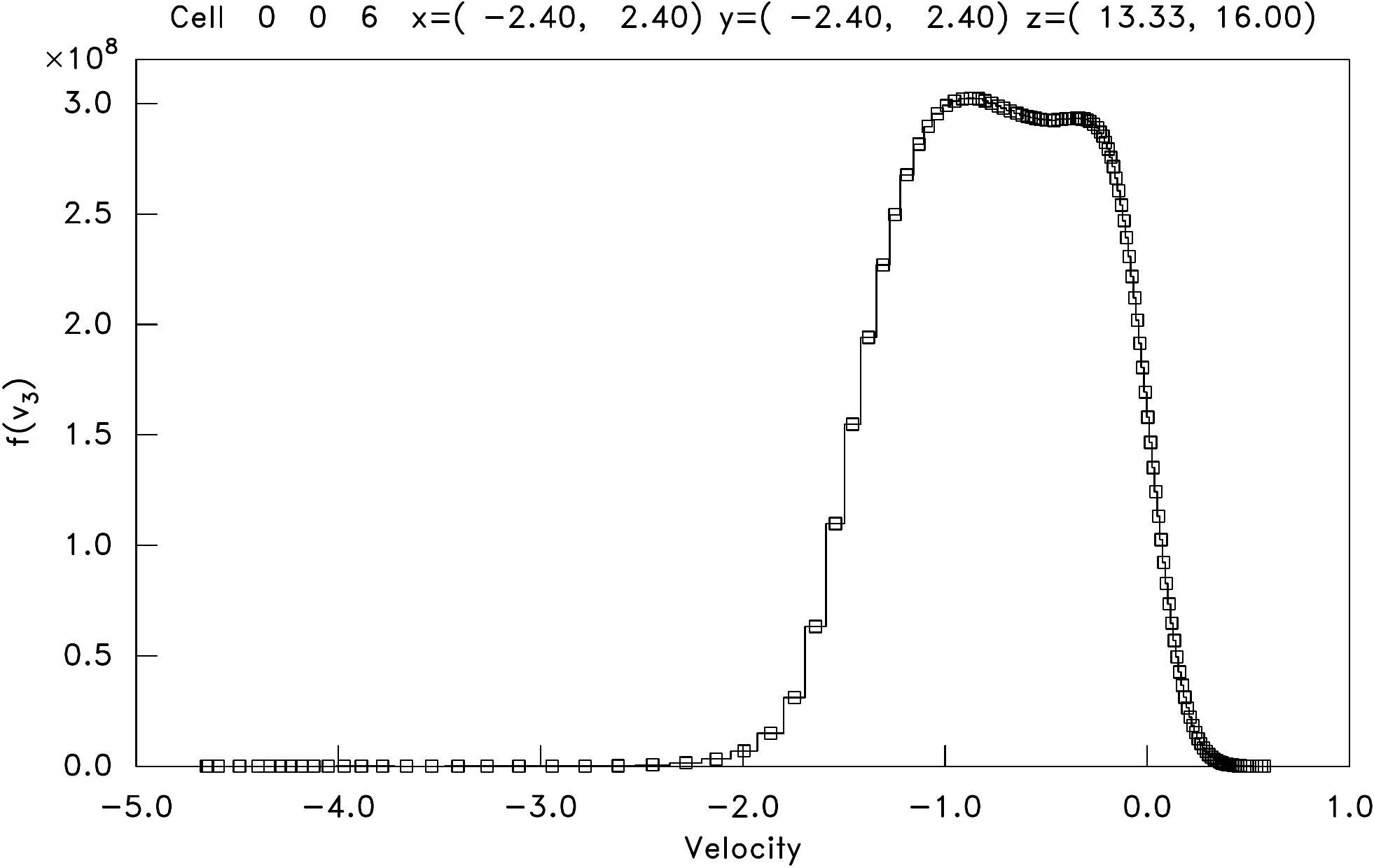}\par\vskip-10pt (b)}

\noindent
\vbox{\noindent
\hsize=0.48\hsize\includegraphics[width=\hsize]{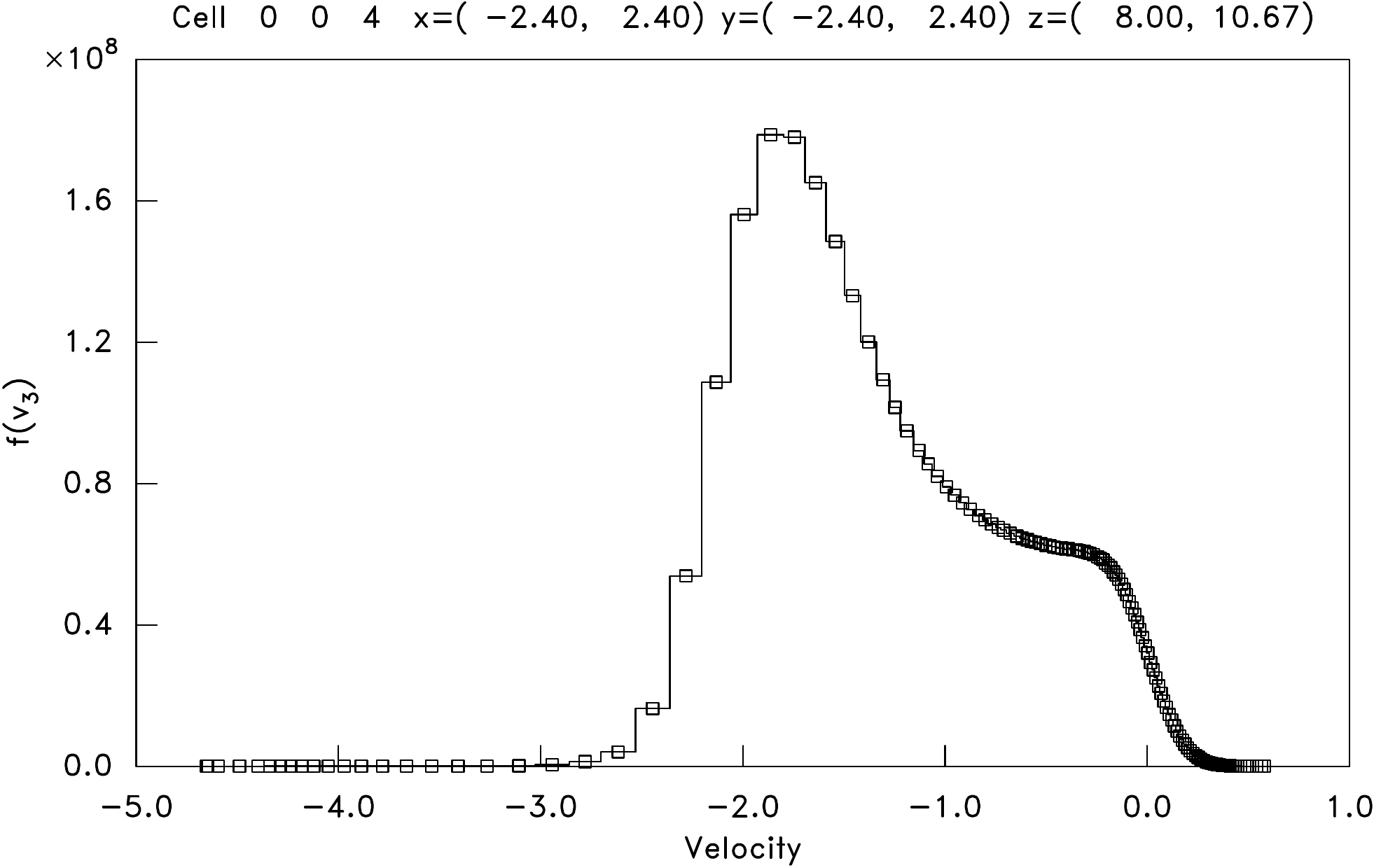}\par\vskip-10pt (c)}
\noindent
\vbox{\noindent
\hsize=0.48\hsize\includegraphics[width=\hsize]{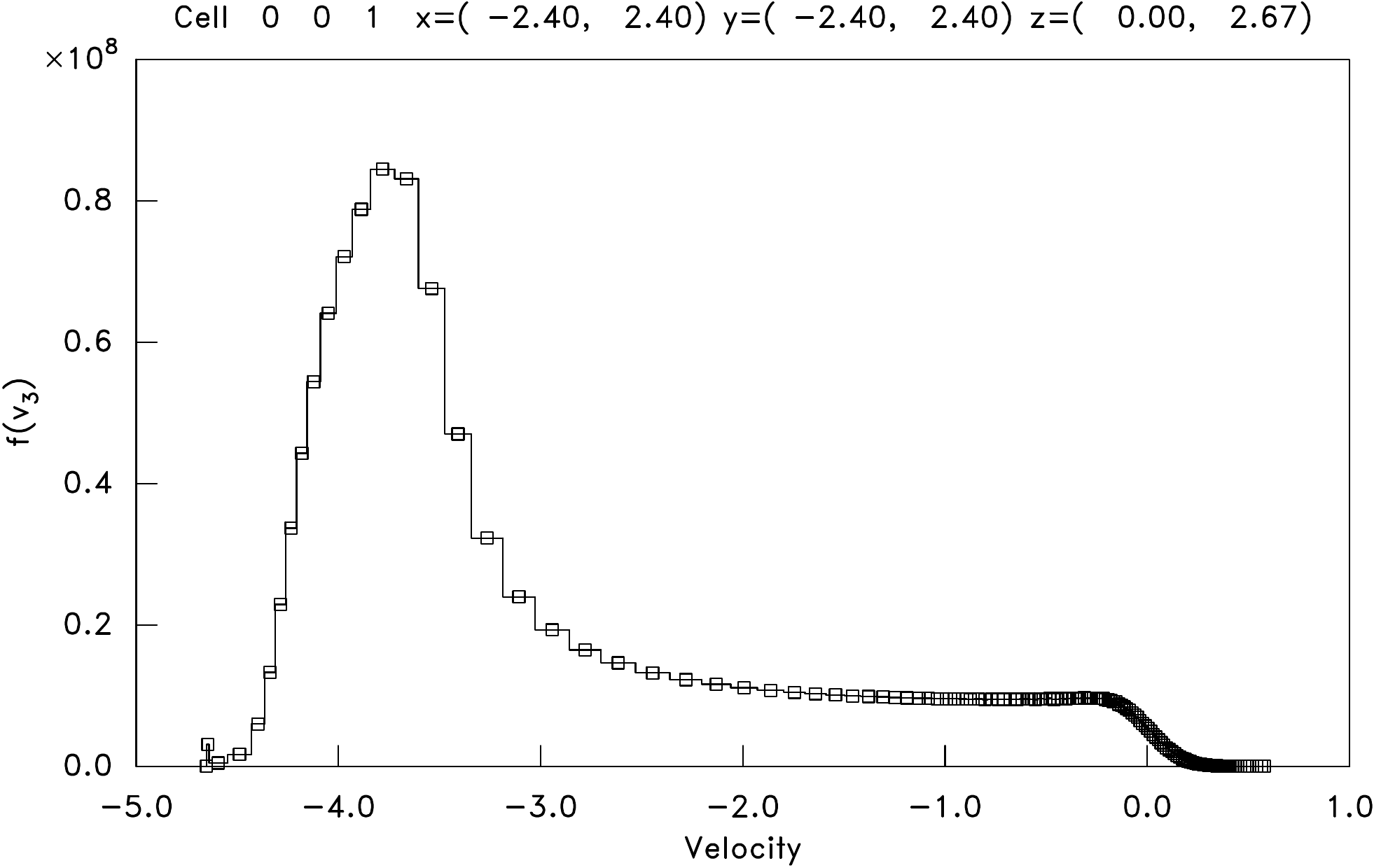}\par\vskip-10pt(d)}
\caption{Ion longitudinal velocity ($v_z$) distribution function at four different
  heights. Collision time $\tau_c=10$.\label{distribsct10}}
\end{figure}

A PIC code gives, in principle, comprehensive information about the
particle velocity distribution function, albeit with limitations on
resolution caused by the statistical noise level. The distribution
function variation with space, for the case of Fig.\
\ref{sheathprofilesct10}, is shown in Fig.\ \ref{distribsct10}.  These
$v_z$-distributions are averaged over transverse velocity and spatial
position, and particles are selected only within the indicated $z$
ranges. We see near the distant domain boundary, a typical drift
distribution (a) whose shape is that of eq.\
(\ref{driftdist}). However, as the sheath edge is approached, strong
distortion of the distribution shape occurs (b). Then passing the
sheath edge the distortion forms itself into an ion beam plus a
trailing plateau in velocity space towards $v_z=0$, which is produced
by collisions. This shape persists, moving deep into the sheath, with
the beam becoming more pronounced and the level of the plateau
decreasing.

\begin{figure}[htp]
  \centering
  \includegraphics[width=10cm]{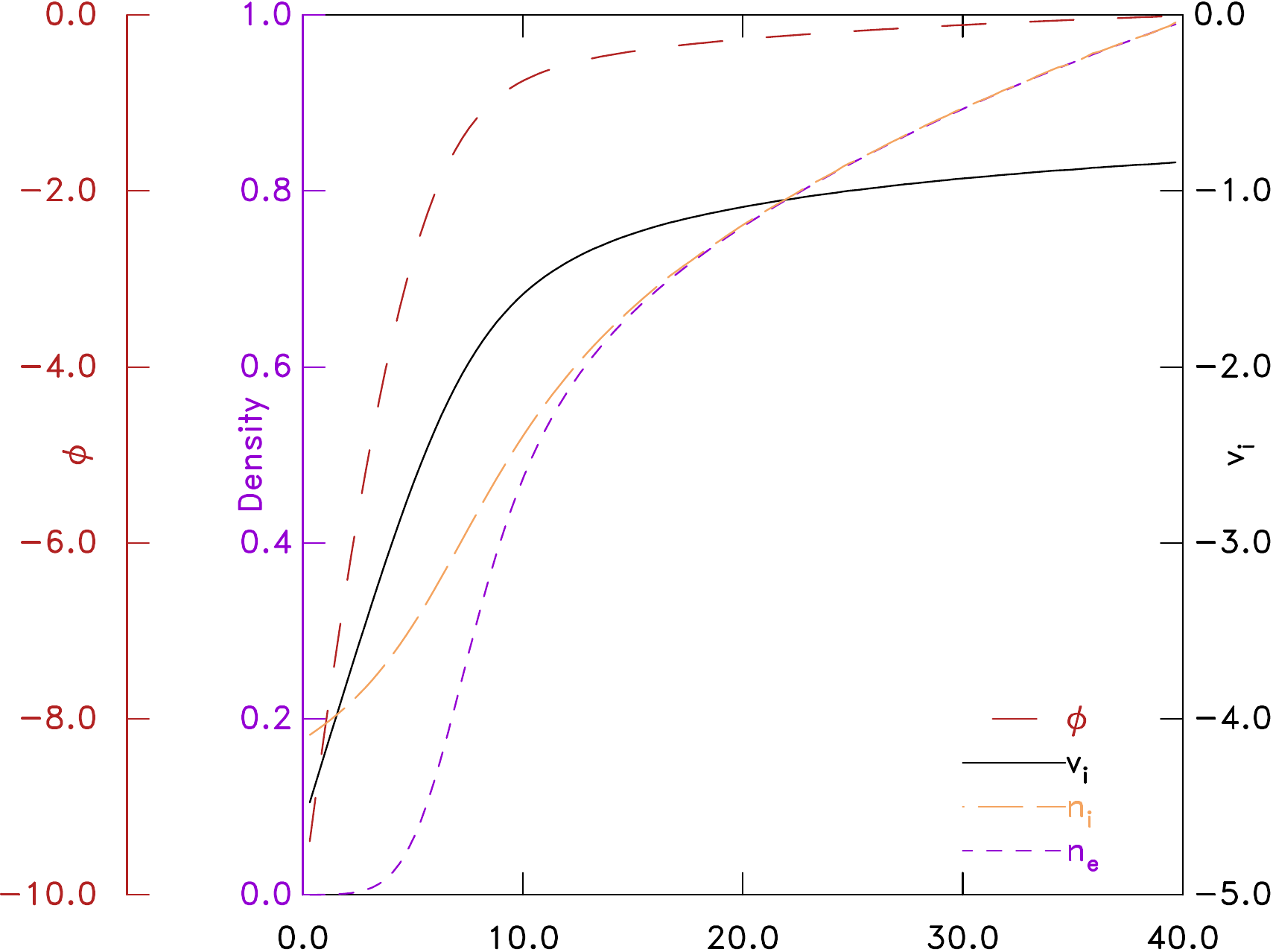}
  \caption{Sheath spatial structure for collision time
    $\tau_c= 100$.}
  \label{sheathprofilesct100}
\end{figure}

\paragraph{Collision time $\tau_c=100$} A much lower collisionality
case is shown in Fig.\ \ref{sheathprofilesct100}.
Its characteristics are similar. However, because the presheath scale,
which is proportional to the mean free path, is ten times longer, it
has a much larger magnitude eigenspeed: $v_{f40}=-0.83$.
 On the
presheath scale, at $z=40$ the solution is already close to the
sheath-edge $z\approx20$, and requires only modest acceleration to
reach the sheath-edge velocity $v_f\approx-1$.

\begin{figure}[htp]
\noindent
\vbox{\noindent\hsize=0.48\hsize
\includegraphics[width=\hsize]{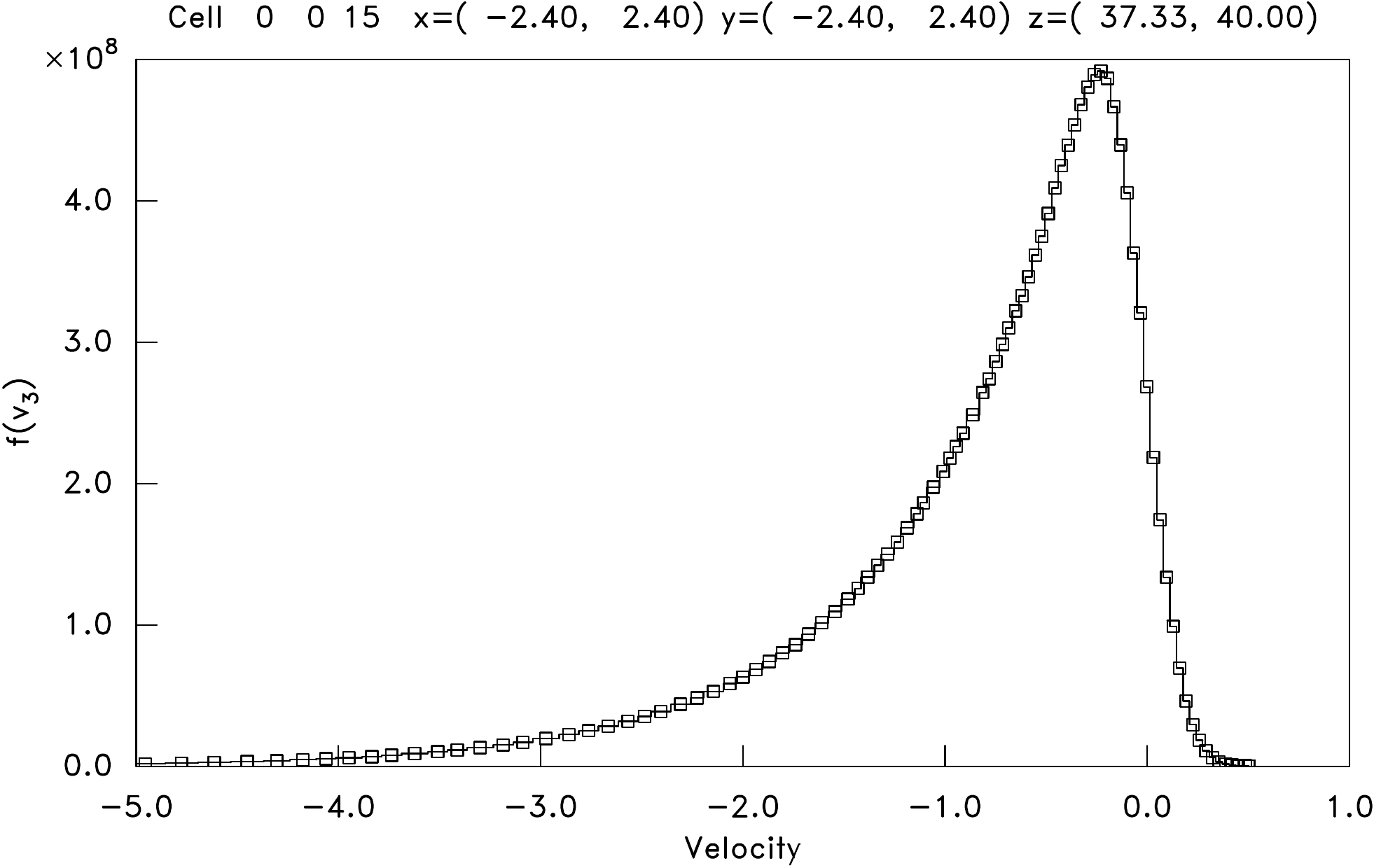}\par\vskip-10pt (a)}
\vbox{\noindent\hsize=0.48\hsize
\includegraphics[width=\hsize]{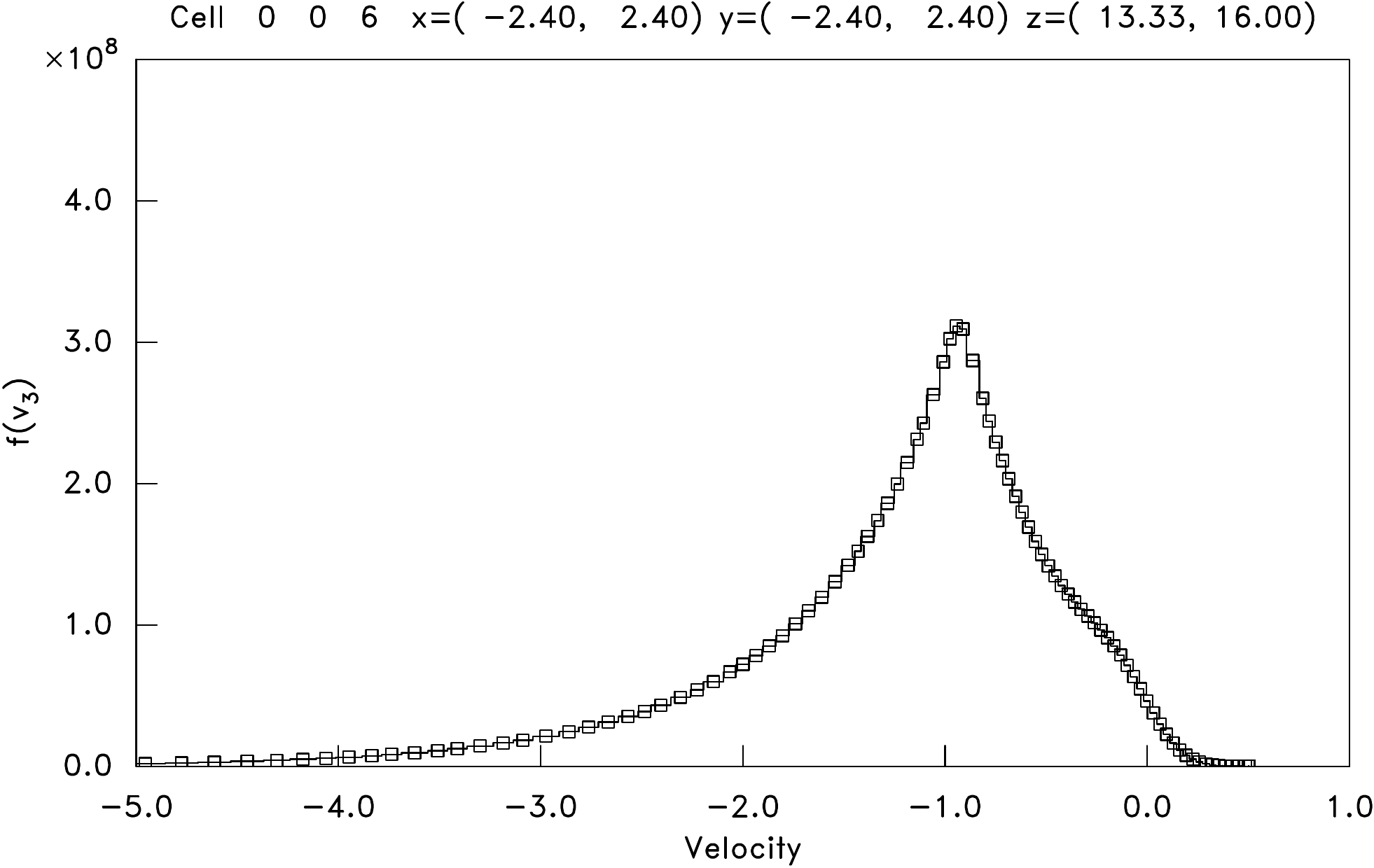}\par\vskip-10pt (b)}

\noindent
\vbox{\noindent\hsize=0.48\hsize
\includegraphics[width=\hsize]{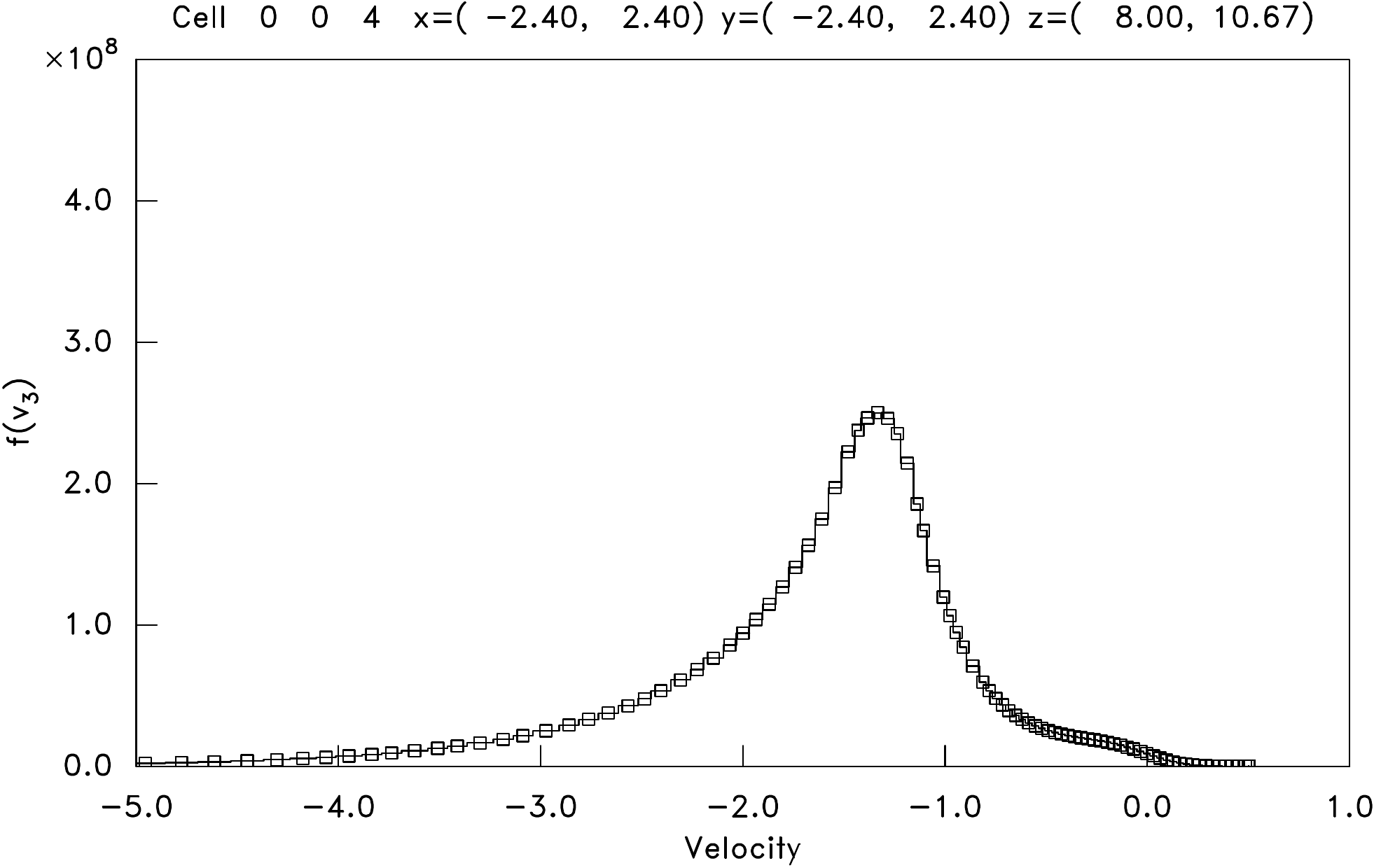}\par\vskip-10pt (c)}
\noindent
\vbox{\noindent\hsize=0.48\hsize
\includegraphics[width=\hsize]{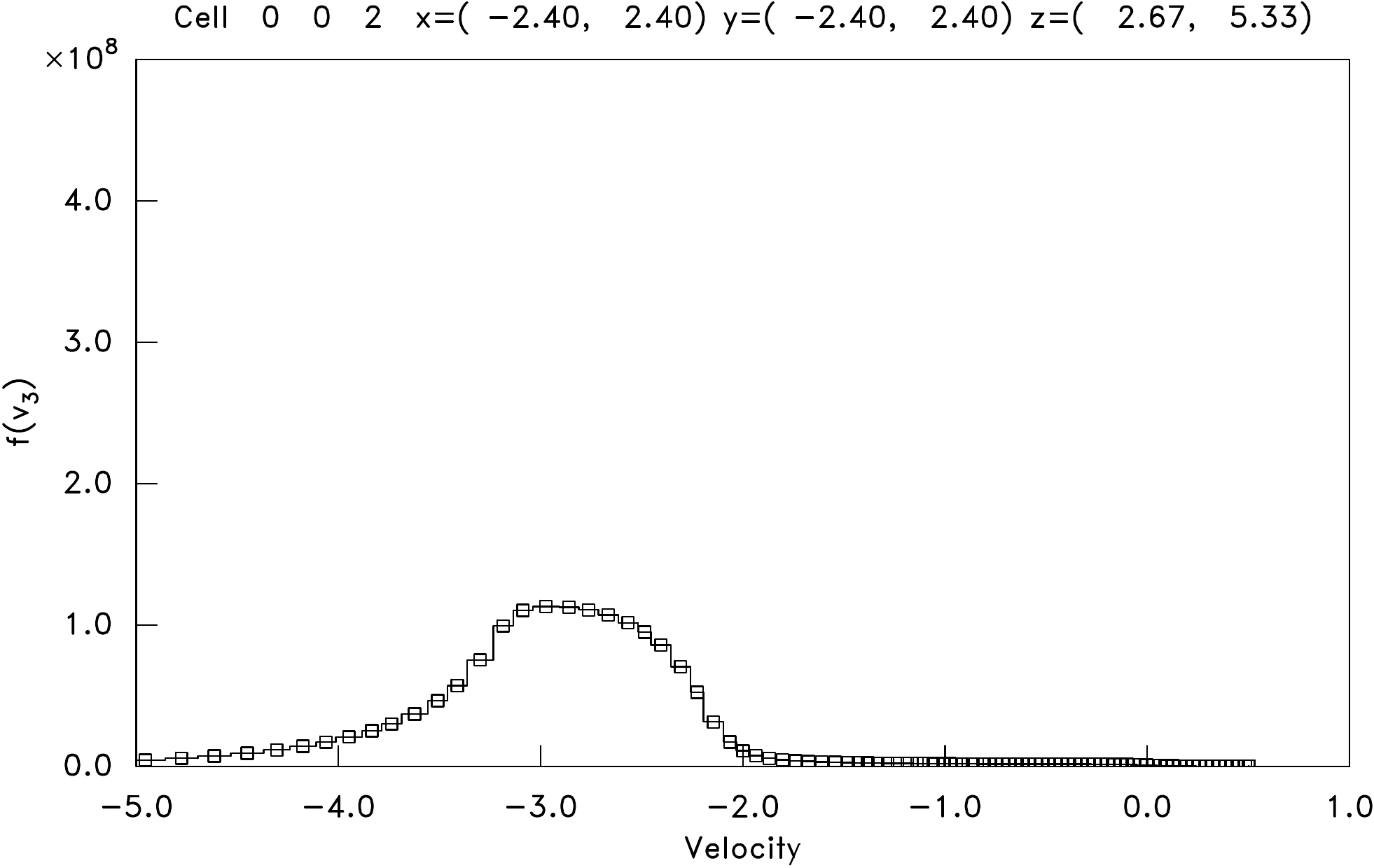}\par\vskip-10pt(d)}
\caption{Ion longitudinal velocity ($v_z$) distribution function at
  four different heights. Collision time
  $\tau_c=100$.\label{distribsct100}}
\end{figure}

The corresponding ion velocity distributions are shown in Fig.\
\ref{distribsct100}. Again, at the entrance to the domain (a) the
distribution is given mostly by the input particle distribution. In
this case where the mean free path is longer than the domain size,
there is some approximation that arises from the presumption that the
particle injection is exactly the drift distribution. As the sheath
edge is approached, (b), acceleration causes the distribution peak to
move to higher speed, then inside the sheath, (c), a separation of the beam
from zero velocity occurs, with only a tiny plateau because collisions
are so infrequent. Deep into the sheath essentially all the ions are
supersonic, the beam is fairly broad, but hardly any ions exist with
speed less than 2.

\begin{figure}[htp]
  \centering
  \includegraphics[width=10cm]{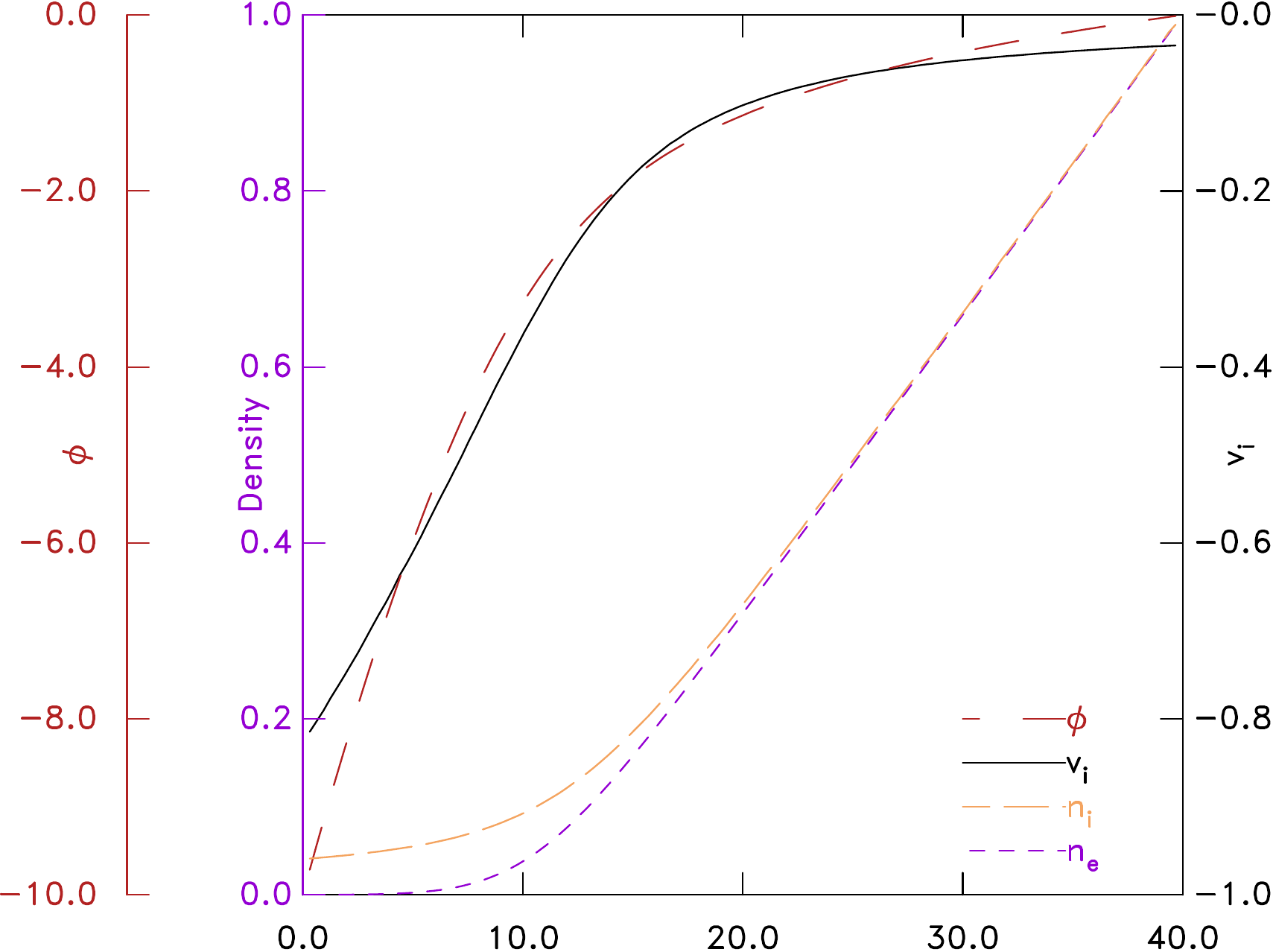}
  \caption{Collsional presheath spatial structure for collision time
    $\tau_c= 1$.}
  \label{sheathprofilesct1}
\end{figure}

\paragraph{Collision time $\tau_c=1$} A much higher collisionality
case is shown in Fig.\ \ref{sheathprofilesct1}. There the
sound-speed mean free path is 1. The eigenspeed is slow
$v_{f40}=-0.035$ and the right hand boundary is deep into the
presheath regime. A large drop of density occurs in the quasineutral
region. It is notable that quasineutrality breaks down at position
$z\approx17$, well before the sound speed is reached. That is a
reflection of there being no longer a clear distinction between sheath
and presheath.

\begin{figure}[htp]
\noindent
\vbox{\noindent\hsize=0.48\hsize
\includegraphics[width=\hsize]{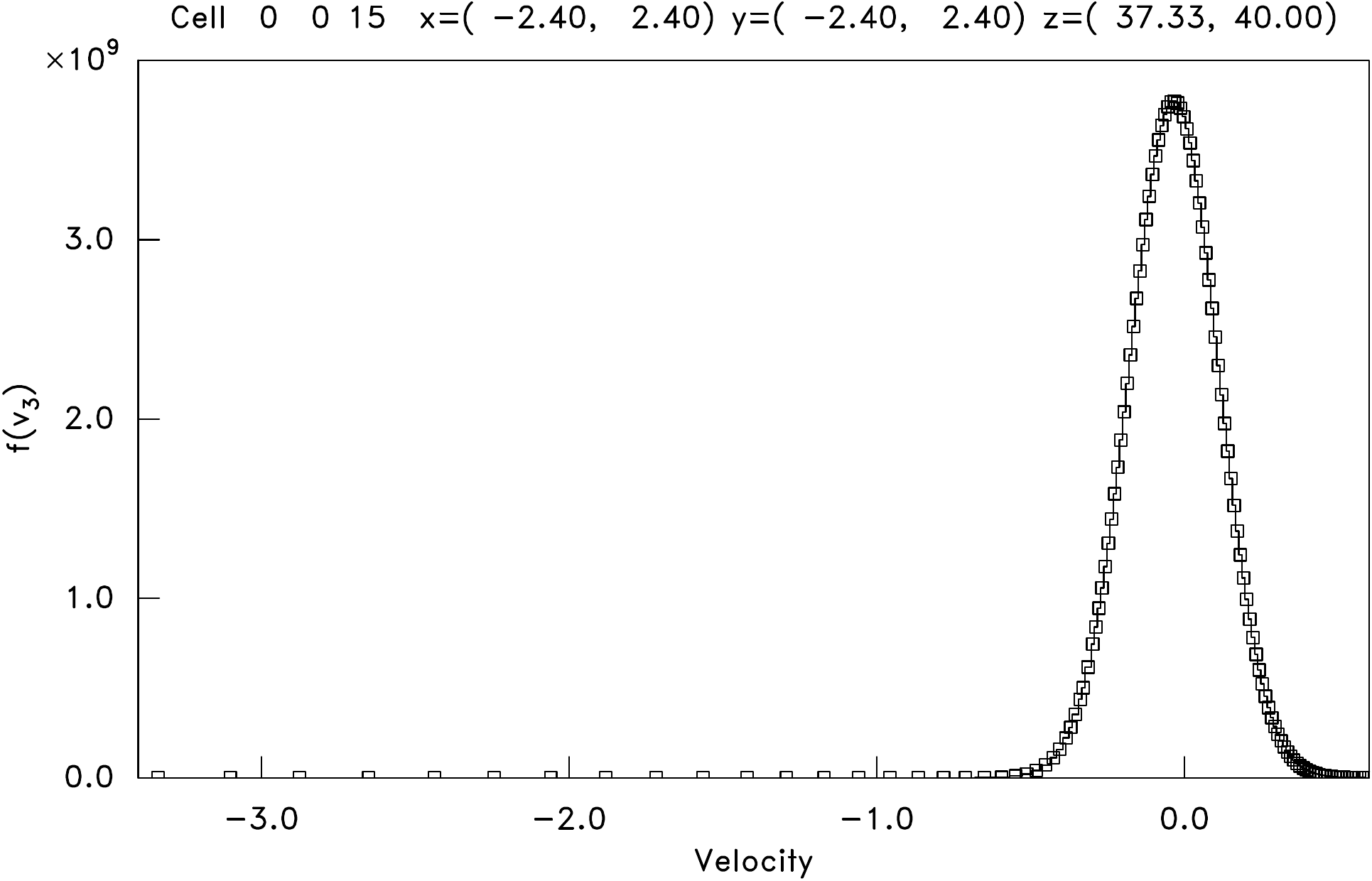}\par\vskip-10pt (a)}
\vbox{\noindent\hsize=0.48\hsize
\includegraphics[width=\hsize]{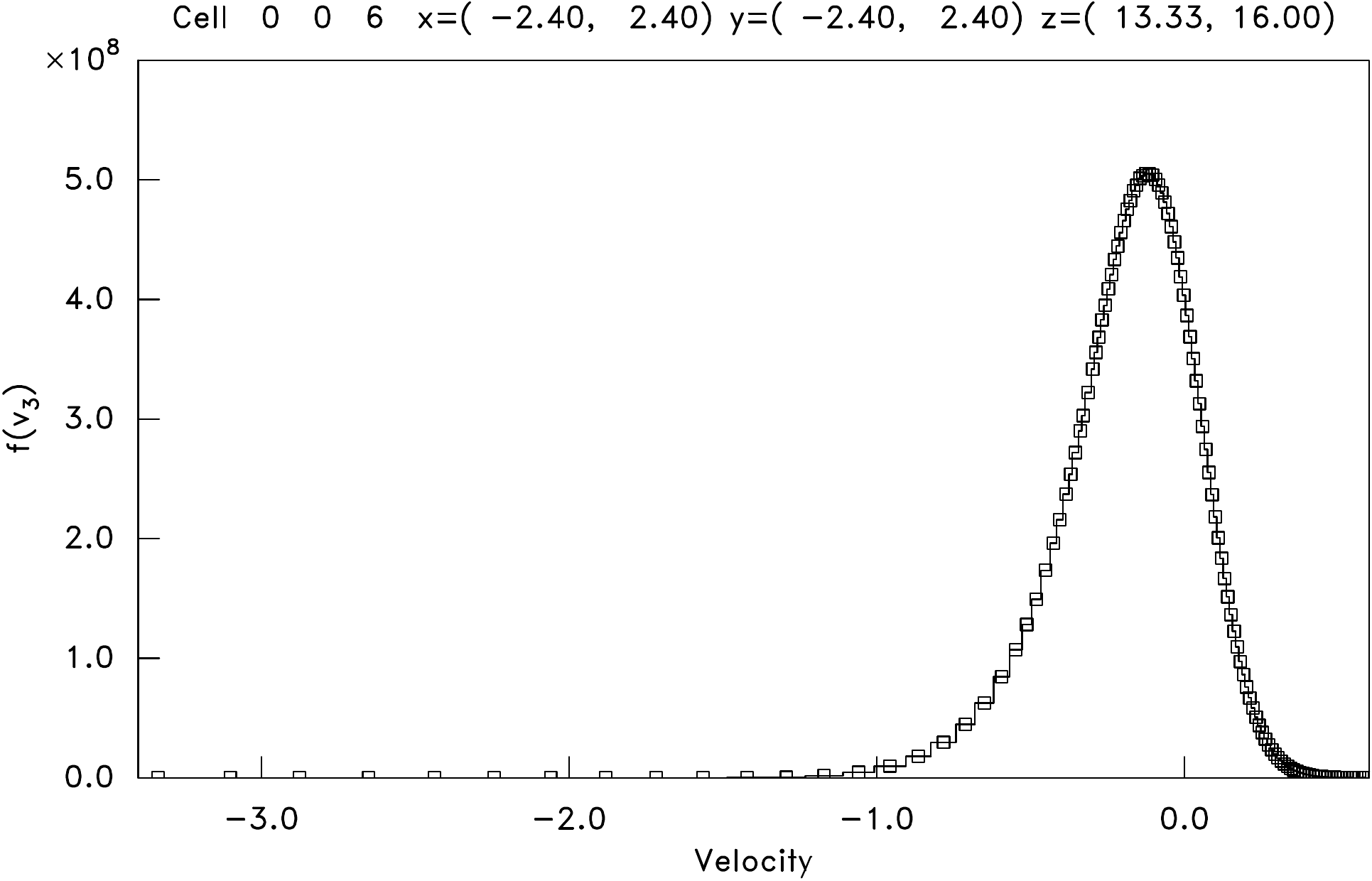}\par\vskip-10pt (b)}

\noindent
\vbox{\noindent\hsize=0.48\hsize
\includegraphics[width=\hsize]{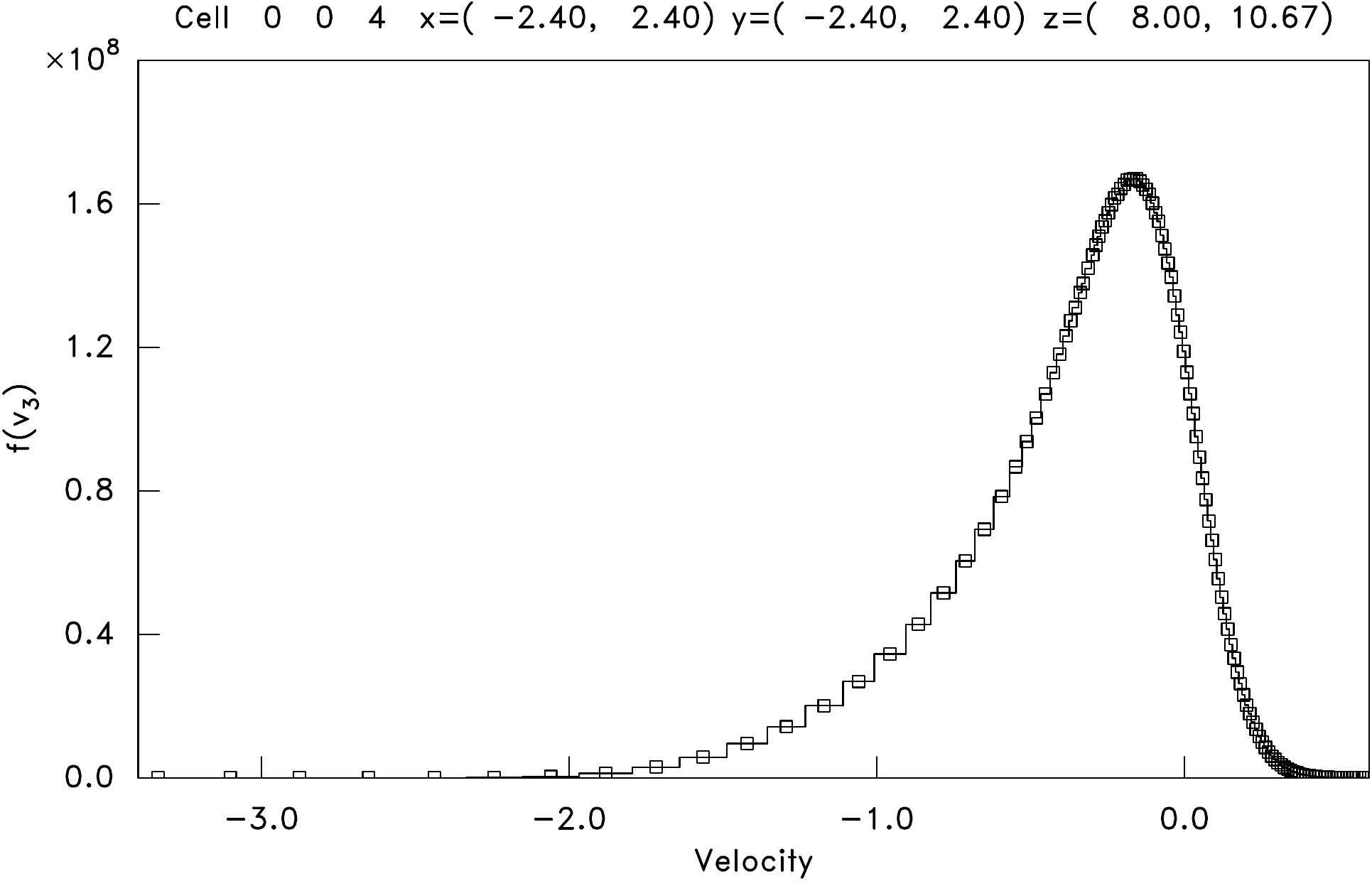}\par\vskip-10pt (c)}
\noindent
\vbox{\noindent\hsize=0.48\hsize
\includegraphics[width=\hsize]{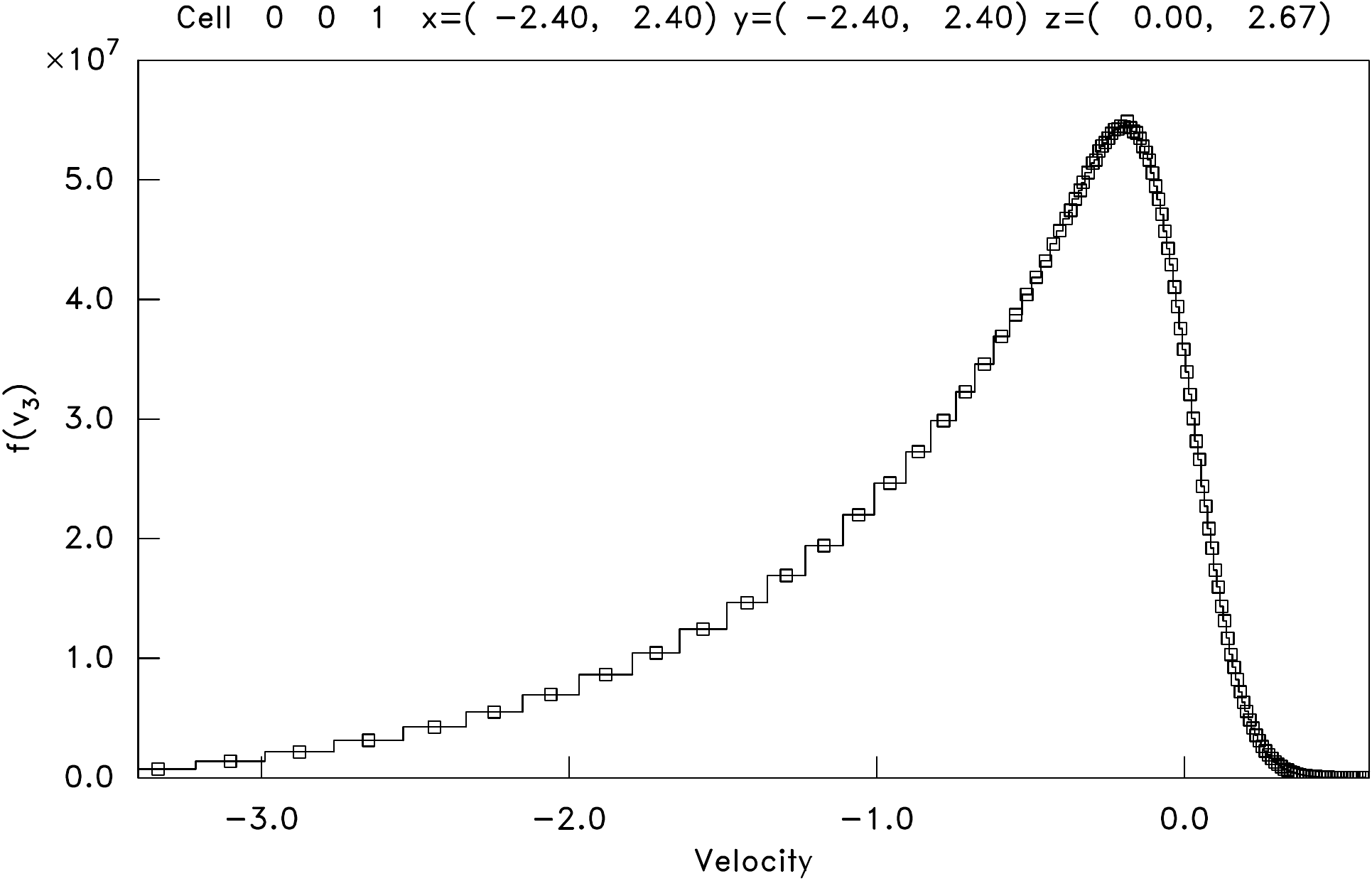}\par\vskip-10pt(d)}
\caption{Ion longitudinal velocity ($v_z$) distribution function at
  four different heights. Collision time
  $\tau_c=1$.\label{distribsct1}}
\end{figure}

The ion distribution functions retain their collisional drift shape
essentially throughout the domain, as shown in Fig.\
\ref{distribsct1}. That is expected since the collision length
is substantially shorter than the sheath thickness.

\section{Grain Force}
\label{grainforce}

In order to calculate the plasma force on a grain located somewhere in
the sheath, COPTIC is run using all the same parameters that have
given the ``bare'' sheath shown in the previous section, but with the
addition of a grain at a chosen height ($z$) and centered in
computational domain in the $x$ and $y$ directions.

The charge on the grain, $Q$, is a key determining factor for the
force. Generally a grain floats at a potential that is minus
approximately 2 to 4 times $T_e/e$ relative to the local
potential. Denote  by $\phi_p$ this potential \emph{difference} between the
grain surface and the local plasma potential. Moreover the
capacitance of grains with radius $r_p$ much smaller than
$\lambda_{De}$ is
\begin{equation}
  \label{capacitance}
  C = Q/\phi_p \approx 4\pi\epsilon_o r_p(1+r_p/\lambda_{s}),
\end{equation}
where $\lambda_s$ is the shielding length.
Consequently, if we define a normalized charge $\bar Q$ as
\begin{equation}
  \label{barq}
  \bar Q = {Qe\over 4\pi\epsilon_0\lambda_{De}T_e},
\end{equation}
then, in normalized units, 
\begin{equation}
  \label{barqnorm}
  \bar Q \approx \phi_p (1+r_p/\lambda_s)r_p .
\end{equation}
The normalized charge $\bar Q$ represents the normalized grain size
times a coefficient $\phi_p(1+r_p/\lambda_s)$ that is between $-2$ and $-4$.

\begin{figure}[htp]
  \centering
\includegraphics[width=8cm]{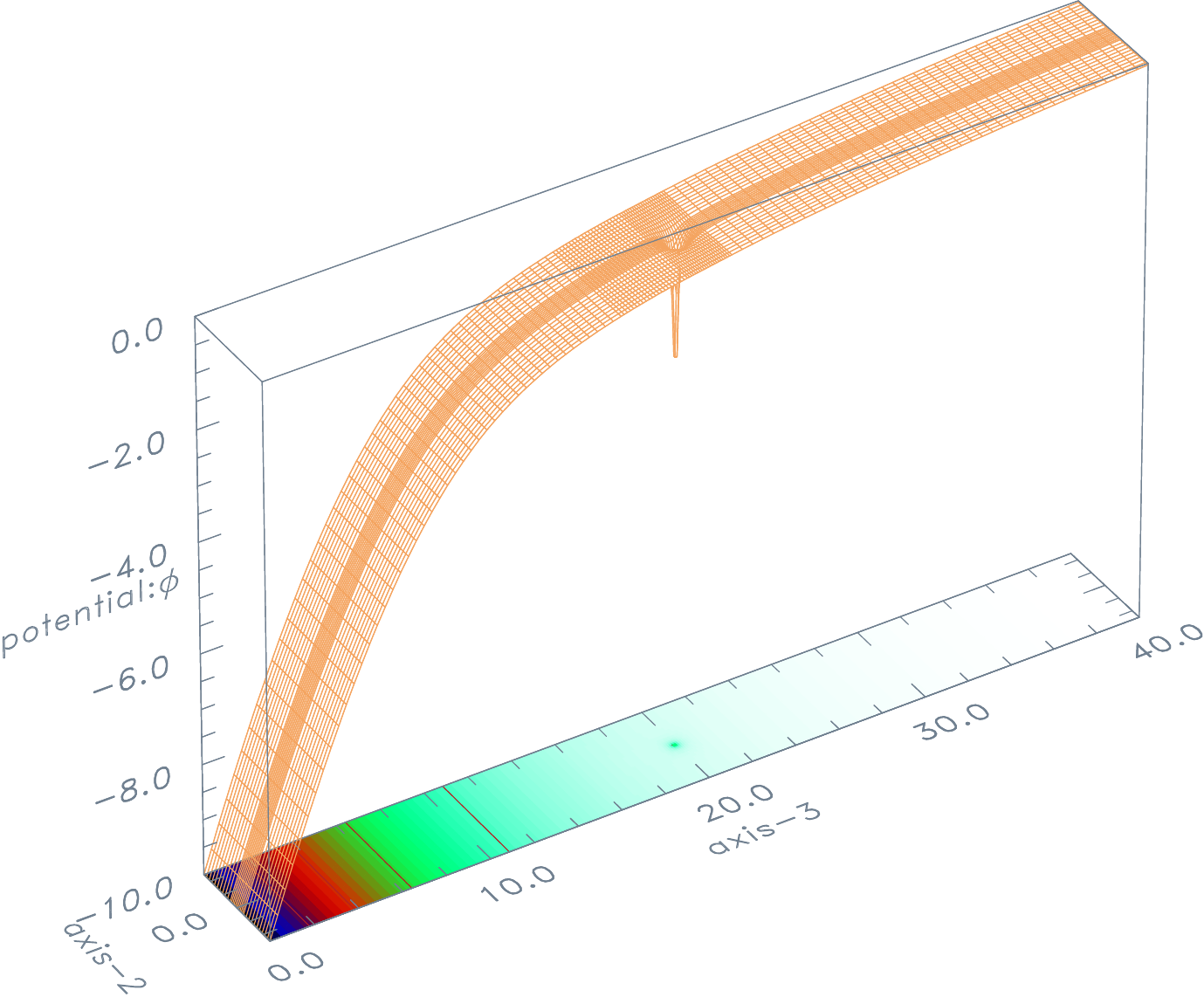}\hskip -3cm
\includegraphics[width=8cm]{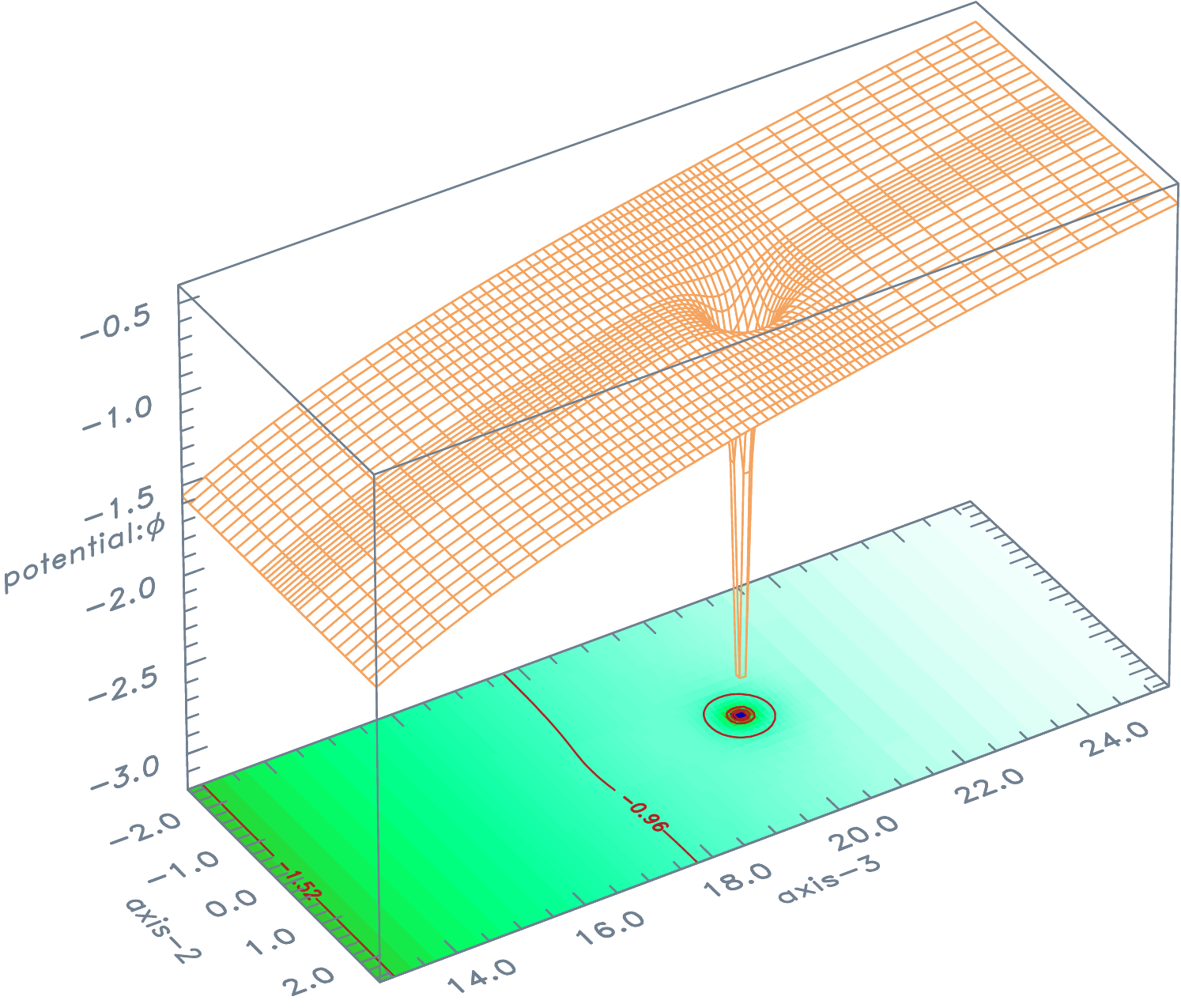}
  \caption{Example of the potential solution when a negatively charged
    grain is present at $z=20$, $|\bar Q|=0.02$, $\tau_c=10$.\label{webz}}
\end{figure}
Fig.\ \ref{webz} illustrates the potential solution in the presence of
a grain. The background sheath solution retains its overall form. The
negative charge (at $z=20$ in this case) causes a substantial
local potential well. It is resolved, as the expanded plot
illustrates, by using a local refinement of the non-uniform mesh, in
the vicinity of the charge. Farther from the charge such high spatial
resolution is not required, so concentrating the cells around the
charge permits the simulation to avoid excessive numbers of cells
overall. When the charge is placed at a different height, naturally
the refinement area is moved appropriately.

Even despite this mesh refinement, a major spatial scale-length
challenge remains.  We are interested in grains that are much smaller
than the Debye length, and simultaneously we are committed to a
computational domain size that is many (40) Debye lengths in length.
It remains too difficult to resolve grain radius as small as, say,
0.05 Debye lengths sufficiently well to represent the sphere. That
would require a mesh spacing of say 0.01 at the grain. The total
domain would then be 4000 times longer than the smallest spacing, and
computational resources would quickly be overwhelmed. So instead, a
variant of the Particle-Particle Particle-Mesh (PPPM)
technique\cite{hockney88} is used.

Within a small ``analytic-radius'' of the point-charge, the electric
potential (and its gradient) is represented partly by an analytic
potential and partly by the potential on the grid. The analytic part
is equivalent to the potential of the point charge shielded by a cloud
of opposite charge whose density is proportional to radius. The cloud
extends out to the analytic-radius where the point-charge is fully
shielded; the analytic field is zero outside the analytic-radius. The
rest of the potential is represented on the grid. It is found by
solving Poisson's equation discretely. Incidentally, the plots of
Fig.\ \ref{webz} are of the \emph{total} equivalent potential, i.e.\
the sum of the discretely represented potential plus the value of the
analytic potential at the grid points.  This PPPM technique means it
is not necessary for the grid to resolve the potential very close to
the grain-center because it is mostly represented there by the
analytic form. Outside of the analytic-radius it is fully represented
on the grid and the analytic form is zero. In this work the
analytic-radius is chosen as 0.4$(\lambda_{De})$. The grid spacing
near the grain is $\sim0.1$.  For each run, $\bar Q$ is specified ---
a fixed grain charge. 

The grain is also surrounded by a ``grain-sphere'', of
radius equal to the grain radius, in which no plasma is present. Ions
that enter the grain-sphere are removed from the simulation. The
radius of the grain-sphere, $r_p$, is chosen in accordance with eq.\
\ref{barqnorm} (ignoring $r_p/\lambda_s$), as
\begin{equation}
  \label{grainsphere}
  r_p= \bar Q/\phi_p = -0.5 \bar Q.
\end{equation}
In other words, the grain size is chosen consistent with grain
surface potential of $-2T_e/e$.  For given $\bar Q$ the precise grain-sphere
radius has only a fractionally smaller effect on total grain force.

The force on the grain is obtained from the final steady solution by
placing around it three reference spheres of radius 0.5, 0.8, and
1.0. The total momentum flux inward across each of the reference
spheres is measured in the code from the sum of Maxwell stress,
electron pressure, and ion momentum
flux\cite{Hutchinson2005,Patacchini2008}.  The total collisional
momentum loss (to neutrals) within each sphere is subtracted, and in
steady state the remainder is the momentum flux to the grain. Any
discrepancies between the grain momentum flux calculated from the
different spheres represents the uncertainty in the calculation, since
an exact calculation should be independent of reference sphere radius.

A code run of 2000 steps starting from the corresponding bare sheath
conditions is generally well converged for the second half of its duration. The
average $z$-force during that period is found. Transverse force
components are observed to be zero to high precision, as expected from symmetry.
Many such runs are carried out to scan different heights and different
normalized charge values.
\begin{figure}[htp]
  \centering
\includegraphics[width=8cm]{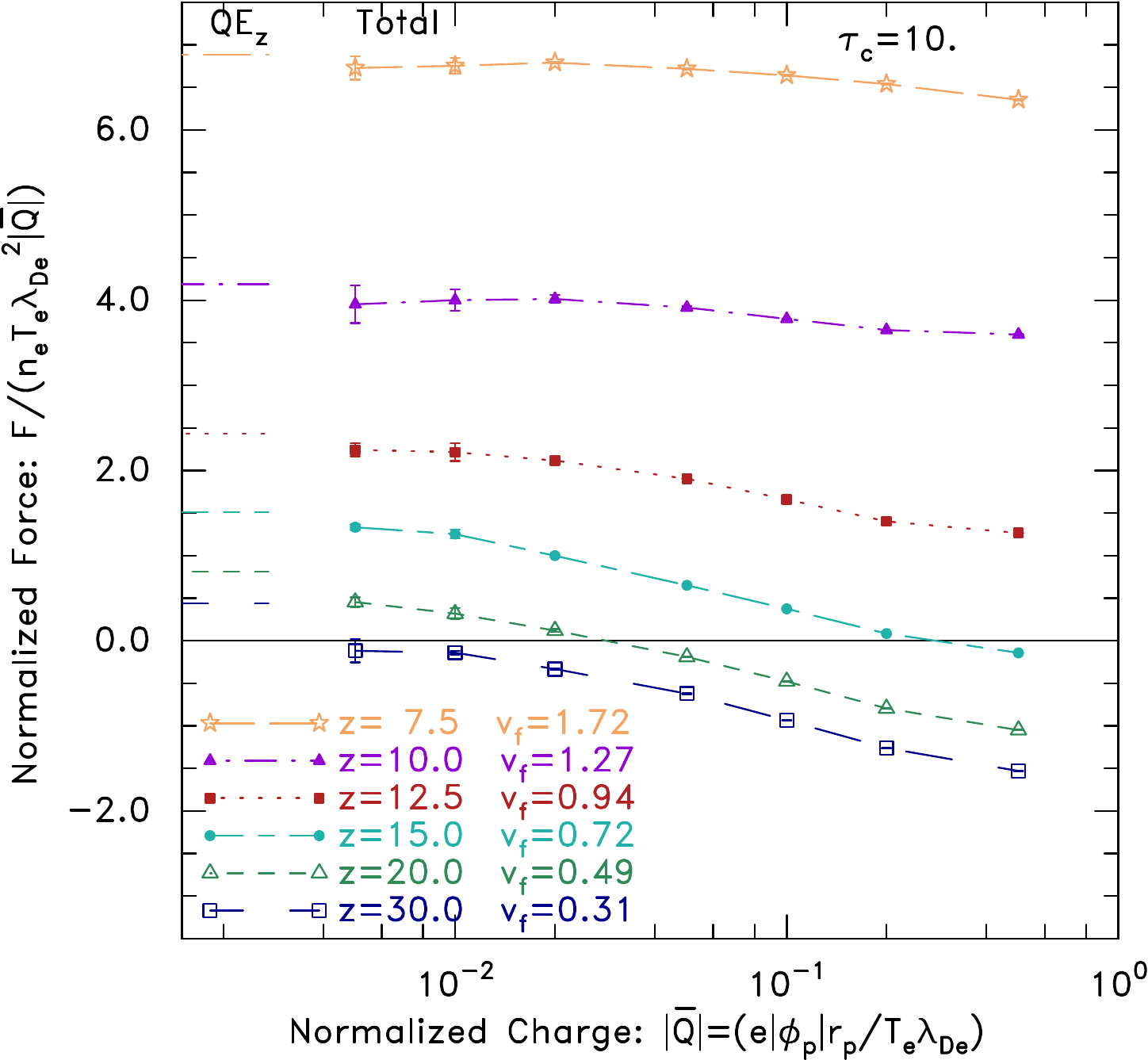}
  \caption{Normalized total plasma force as a function of normalized charge, for various different heights in a sheath with collision time $\tau_c=10$.\label{ct10}}
\end{figure}

Fig.\ \ref{ct10} shows the results for the moderate collisionality
case $\tau_c=10$. The force measured and plotted is the total plasma force, which
includes the effect of the background electric potential gradient
acting on the charge, as well as the ion drag force arising from the
plasma flow past the charge. It is in units of $n_eT_e\lambda_{De}^2$,
but normalized by dividing by $|\bar Q|$. The background electric
field force ($-Q\nabla\phi$ using the bare sheath potential),
normalized in this way, is therefore independent of $|\bar Q|$. Its
value is indicated by short horizontal lines at the left of the plot,
separated by a gap from the full plasma force data.  In a
collisionless situation, the normalized ion drag force at small $|\bar
Q|$ is proportional to $|\bar Q|$. Therefore one expects the limit of
the total force at small $|\bar Q|$ to equal simply the background
electric field force. We do not actually proceed to the limit, because
the force becomes increasingly difficult to measure and subject to
uncertainties at low $|\bar Q|$ . However, the plot appears to confirm
the expectation.

One way to use the data of this plot is to suppose that the non-plasma
forces (e.g.\ gravity) are known; and then find the place where a
grain of a certain charge is in equilibrium with total force equal
zero. If non-plasma forces are zero, for example, then equilibrium is
at Force per charge = 0. A grain at $z=20$ is then in equilibrium if it
has a charge of just under $|\bar Q|=0.05$; a grain of charge $|\bar
Q|=0.2$ will float at $z\approx15$; and so on.

\begin{figure}[htp]
\noindent
\vbox{\noindent\hsize=0.48\hsize
\includegraphics[width=8cm]{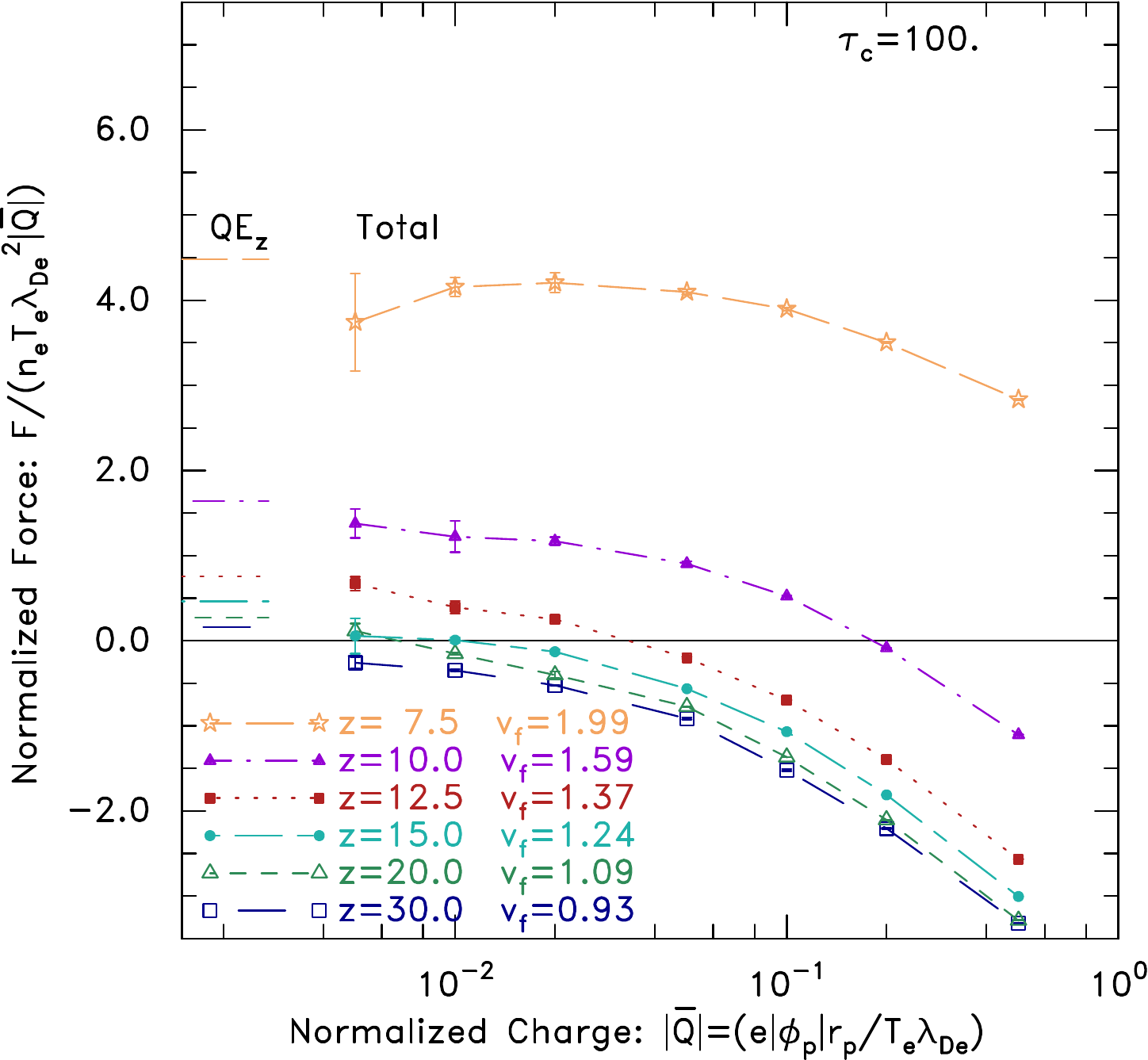}\par\vskip-10pt\quad (a)}
\vbox{\noindent\hsize=0.48\hsize
\includegraphics[width=8cm]{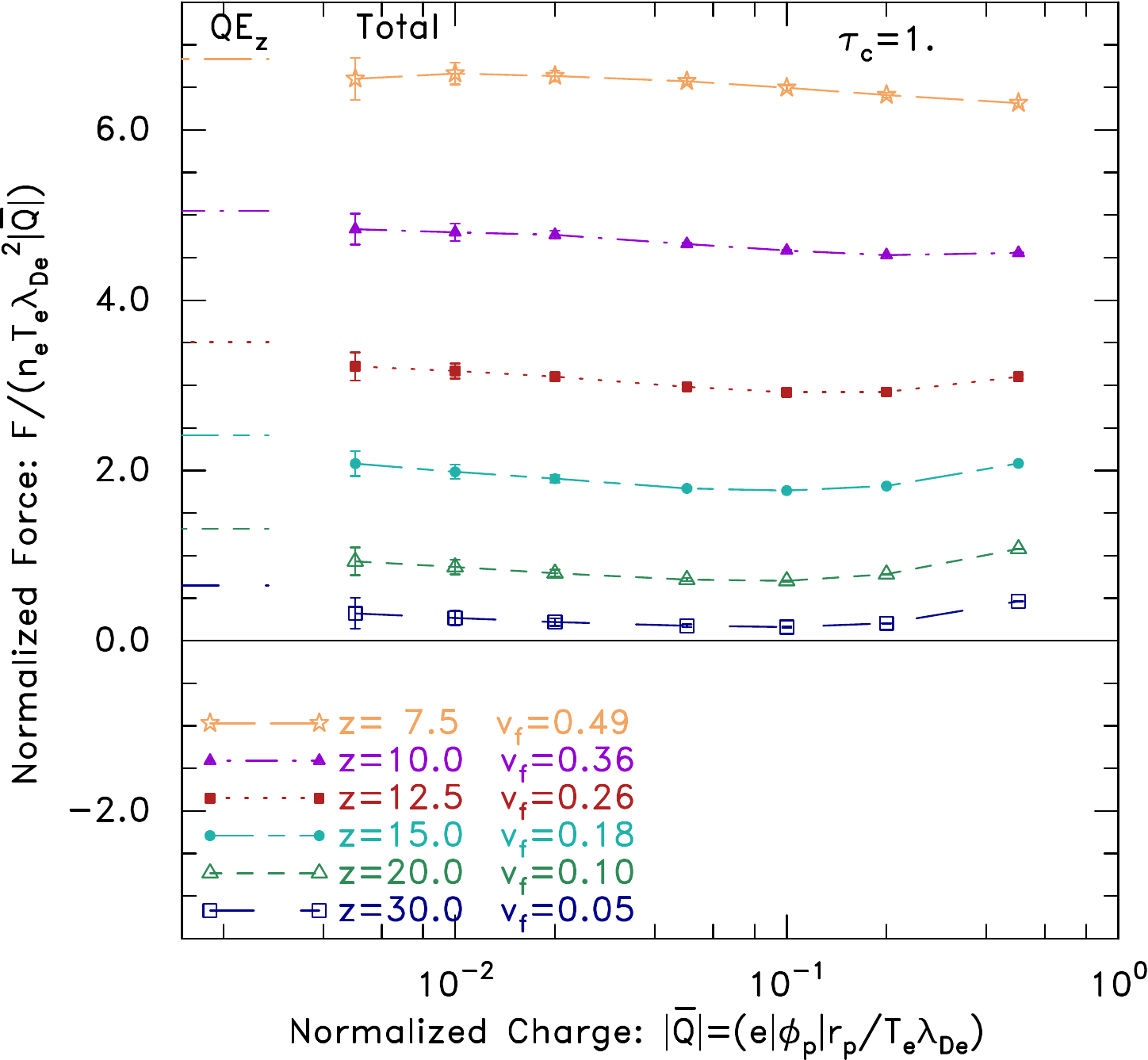} \par\vskip-10pt\quad (b)}
\caption{Normalized force for collision times (a) $\tau_c=100$ and (b)
  $\tau_c=1$.}
  \label{ct1}
\end{figure}
Fig.\ \ref{ct1} shows force results for the low, (a) $\tau_c=100$,
and high, (b) $\tau_c=1$, collisionality sheaths. As before the force
is mostly electric field force at the left hand, low $|\bar Q|$, side
of the plot. The predominant differences are due to the altered
potential and velocity profiles. It is notable that the variation of
force with $|\bar Q|$ is substantially less for the
high-collisionality sheath and the force is positive even above
$z=20$. If non-plasma forces were negligible for that case, a grain
would float high above the sheath at a position where the flow
speed is smaller than $0.1c_s$ (see Fig.\ \ref{sheathprofilesct1}). One should be cautious in situations
with strong collisionality about making the naive assumption that the
flow at the grain is necessarily sonic. By contrast, at low
collisionality, $\tau_c=100$, even gravitationless grains don't float
above $z=20$, and the speed is indeed essentially equal to $c_s$ or
greater (see Fig.\ \ref{sheathprofilesct100}) at that height or below.

\section{Comparison with Uniform-Plasma Force}
\label{comparisonuniform}

The observed force values are now compared in detail with the force that
would be predicted for a uniform-plasma with appropriate conditions.
\begin{figure}[htp]
  \centering
  \includegraphics[width=10cm]{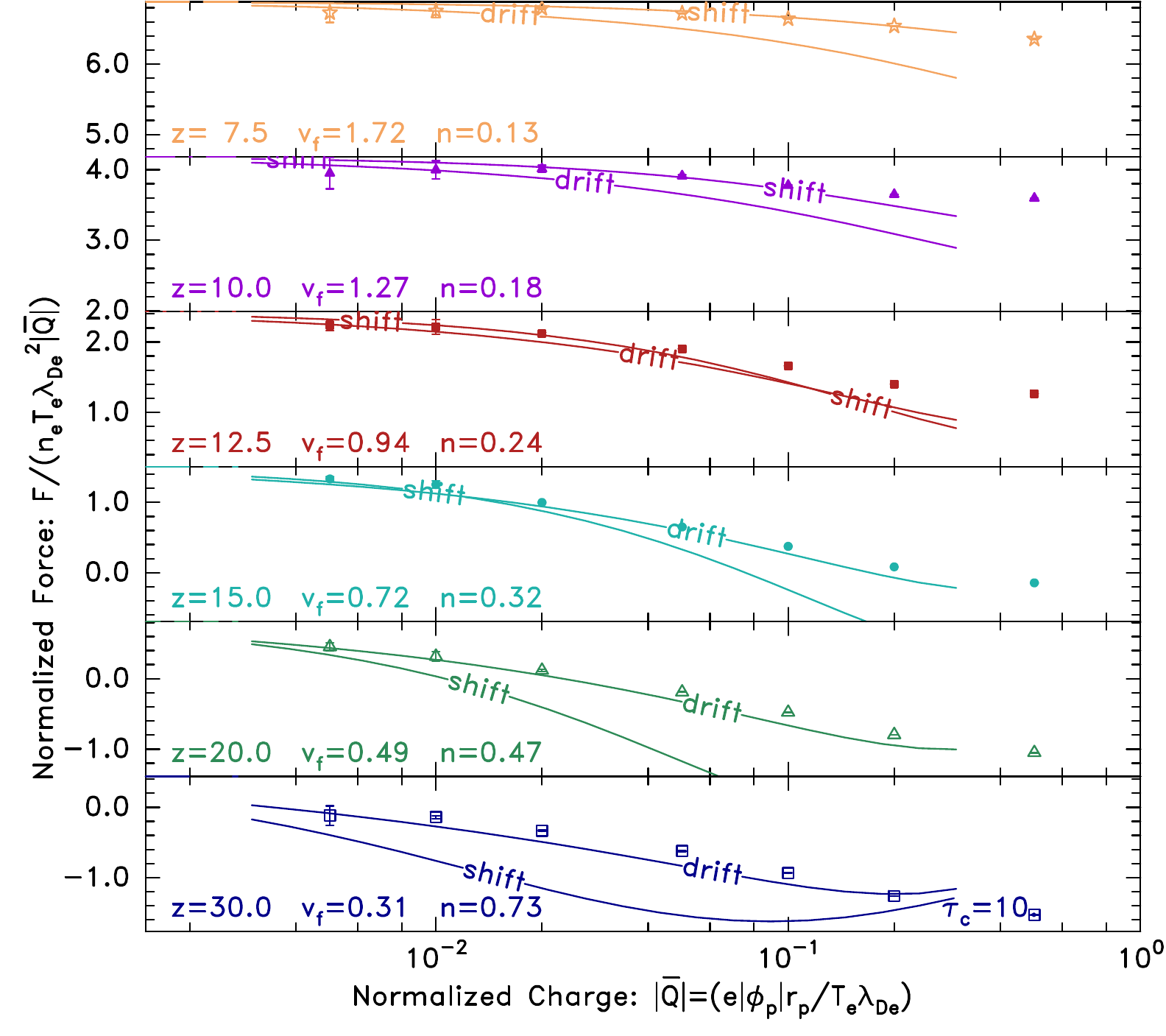}
  \caption{Comparison of force derived from COPTIC with a combination
    of the background electric field force and the fitted ion drag
    force\cite{Hutchinson2013a}. Moderate collisionality
    $\tau_c=10$. The solid lines are theory for a \emph{shift}ed
    Maxwellian and the \emph{drift} distribution [eq.\ \ref{driftdist}].}
  \label{dragcomp10}
\end{figure}

For each position $z$ the value of the density $n_i$, flow velocity
$v_f$, and electric field $-\nabla\phi$ are obtained from the data of
section \ref{colnsheath}, i.e.\ the bare sheath, for the appropriate
value of the collisionality. The ion drag force for a uniform plasma
with that density and velocity is obtained from the fitted analytic
approximations to computational evaluations of the drag force,
recently published\cite{Hutchinson2013a}. Details are summarized in
the appendix. The value of the Debye length used in the
analytic force is equal to the reference value (1) divided by the
square-root of the local normalized density (at $z$, relative to the reference
density at $z=40$); and the analytic force is multiplied by the local normalized
density to render it into the units normalized to the reference
density. 

The analytic drag force (which is negative) is added to the (positive)
electric field force given by the grain charge multiplied by the
background electric field. The result is the predicted total
uniform-plasma force for a grain residing in a plasma whose properties
are those of a bare sheath at the position of the grain. However,
notice that the electric field is \emph{not} exactly the same as it
would be in a uniform plasma. In a plasma in which the ion flow arises
from a flow of background neutrals --- the ``shift'' case with shifted
Maxwellian ion distribution --- there would be zero electric field. In
a uniform plasma where the ion flow is driven by an electric field,
the distribution takes the ``drift'' form and the electric field is
directly related to the flow velocity through $-\nabla \phi =
v_f/\tau_c$ (in normalized units).  In neither case is the
uniform-plasma electric field exactly equal to that in the bare
sheath, because even in the drift case, ion acceleration in the sheath
is part of the net average force balance, causing the bare-sheath
electric field to deviate from $v_f/\tau_c$. This difference in
electric field means that the uniform-plasma ion drag prediction
cannot be expected to be exactly equal to the drag in a non-uniform
sheath.

In Fig.\ \ref{dragcomp10} are shown the COPTIC data compared with the
analytic predictions, for the moderate collisionality case. The upper
boundary of each frame is placed at the value of the electric field force. The
analytic fits for drift distributions are not validated for
$r_p/\lambda_{De} > 0.1$, which corresponds to $|\bar{Q}|>0.2$, so the
fit lines are not extended to the maximum COPTIC point at
$|\bar{Q}|>0.5$. Two fit lines are shown, corresponding respectively
to the drift or shift distributions (with specified $v_f$). We observe
that for distant points such as $z=$ 30, 20, 15, the \emph{drift} curve fits
the COPTIC data rather well, whereas the \emph{shift} curve does not. The
observed distribution function in this height range is much better
approximated by the drift distribution, as seen in
Fig. \ref{distribsct10}. However, as the height decreases $z=$ 10, 7.5,
into the sheath the ions accelerate so as to adopt a more
``beam-like'' distribution (Fig.\ \ref{distribsct10}c). The
corresponding force then agrees better with the shift analytic expression.

In all cases, as the charge becomes small, the values appear to be
approaching the electric-field value. However, the force uncertainty
estimated from the difference between the values measured by different
spheres, eventually becomes significant. Recall that the plot is of
the force divided by $\bar Q$, so absolute uncertainty is magnified at
low $|\bar Q|$ in the normalized force. We therefore do not explore
below $|\bar Q|=0.005$.

\begin{figure}[htp]
  \centering
  \includegraphics[width=10cm]{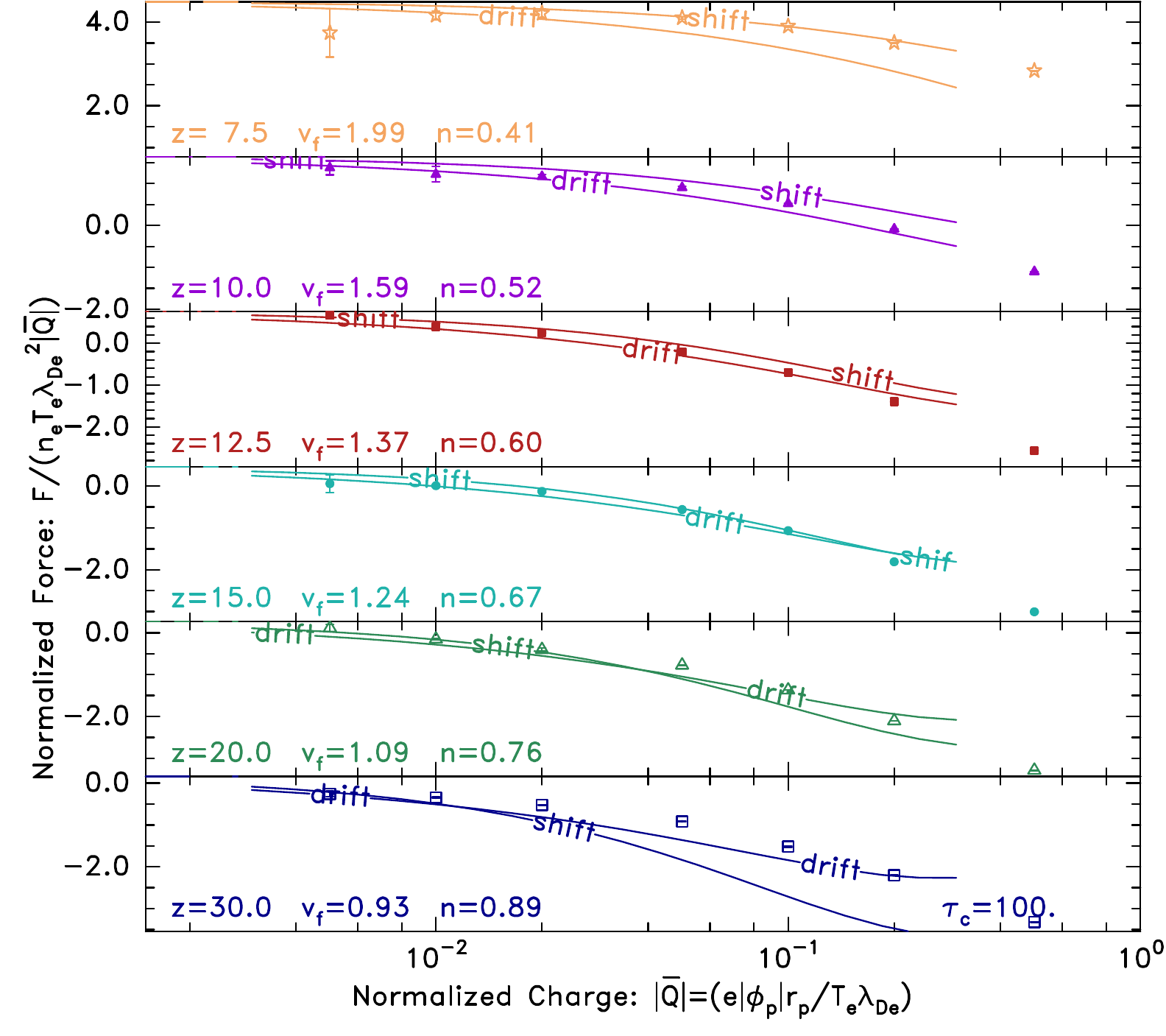}
  \caption{As for Fig.\ \ref{dragcomp10} except for
 low collisionality $\tau_c=100$.}
  \label{dragcomp100}
\end{figure}

\begin{figure}[htp]
  \centering
  \includegraphics[width=10cm]{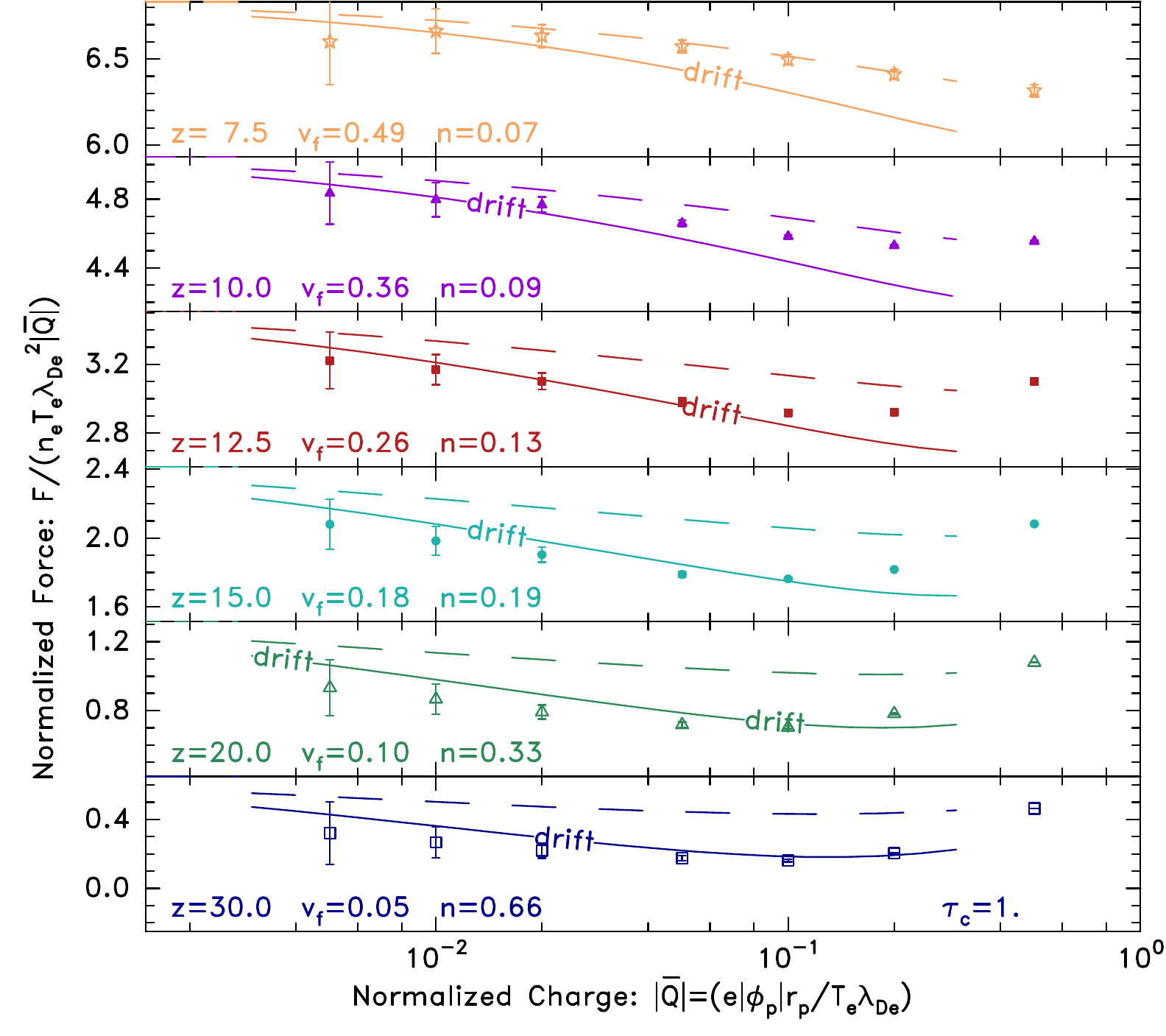}
  \caption{Comparison of force derived from COPTIC with a combination
    of the background electric field force and the ion drag
    force including (solid line) or excluding (dashed line) the
    collisional force factor\cite{Hutchinson2013a}. High collisionality $\tau_c=1$.}
  \label{dragcomp1}
\end{figure}

Fig.\ \ref{dragcomp100} shows the same comparison for the low
collisionality sheath $\tau_c=100$. It shows much the same trend of a
transition between drift- and shift-like distributions and
forces; although the predicted differences between the forces are
somewhat smaller in this case.

Fig.\ \ref{dragcomp1} shows the comparison for the high-collisionality
case, $\tau_c=1$. We know from Fig.\ \ref{distribsct1} that for this
case the bare sheath shows drift-like distributions throughout the
range of $z$. Accordingly, the drift-force prediction agrees
substantially better than the shift-force, at all positions. No
shift-force predictions are plotted. Instead, in addition to the full
drift-force prediction, the low-collisionality drift-distribution
force is shown as a dashed line. Its drag component is approximately a
factor of two less than the full prediction because as has been
documented\cite{Hutchinson2013a} there is a factor of two enhancement
of the drift force directly by collisions at collisionalities like
this. We therefore see that it is necessary to include the collisional
force enhancement in order to obtain agreement with COPTIC's observed
forces.

\begin{figure}[htp]
  \centering
(a)\includegraphics[width=6cm]{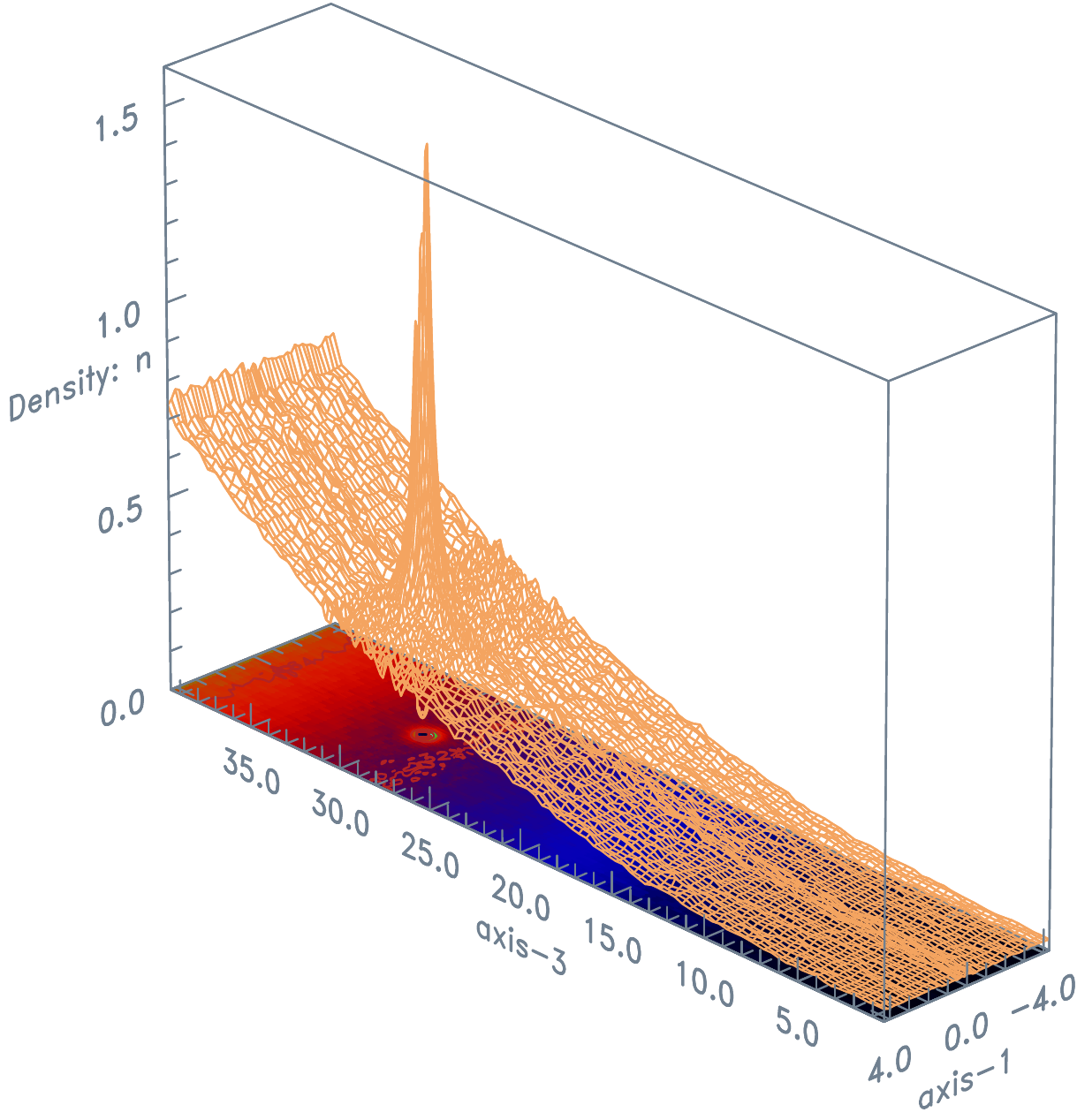}\hskip -2.2cm
\includegraphics[width=8cm]{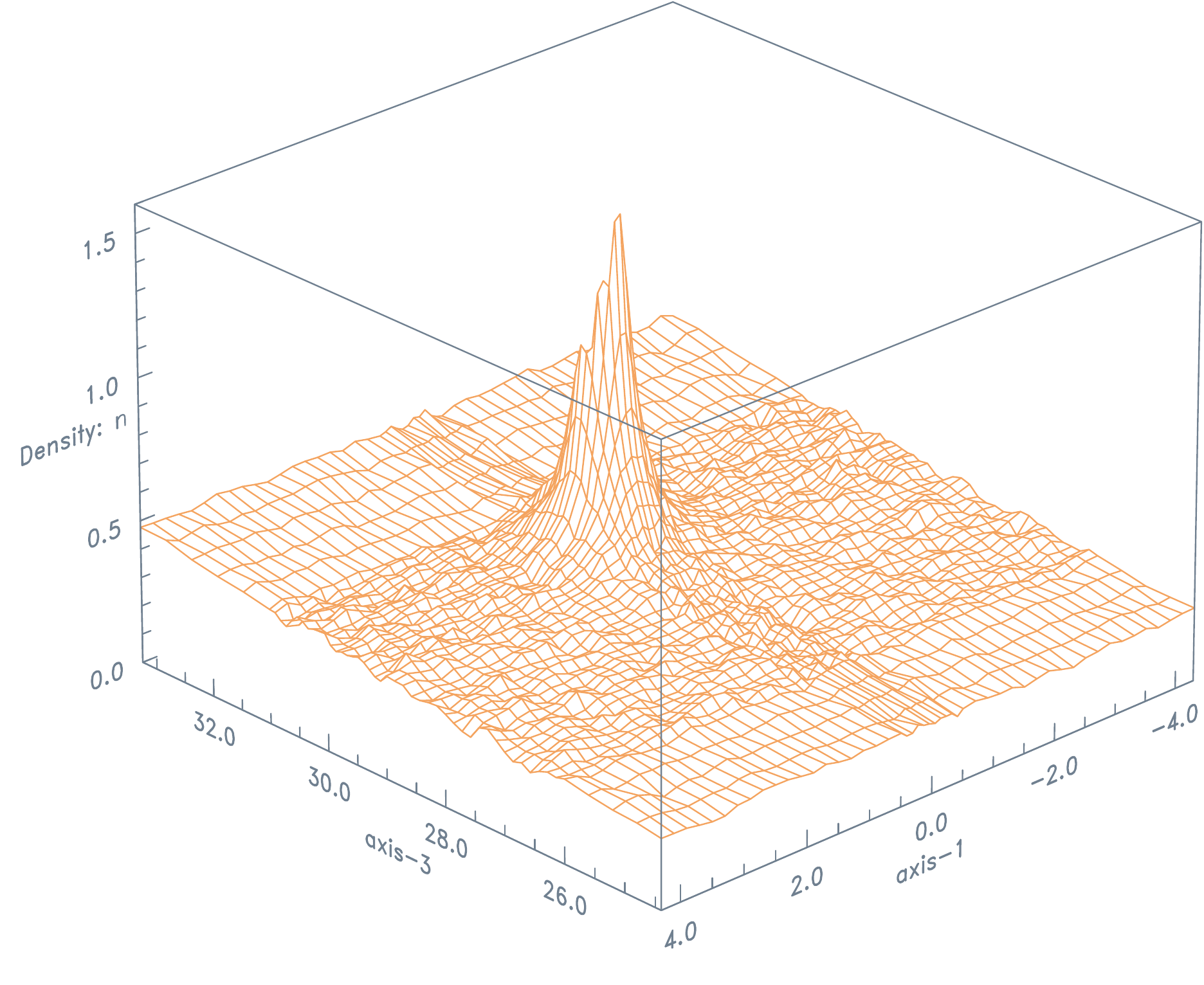}(b)
  \caption{Illustration of the strong perturbation of the sheath
    density by a large charge. $|\bar Q|=0.05$, in the high
    collisionality case $\tau_c=1$. (a) Entire domain. (b) Vicinity of
  charge.}
  \label{denweb}
\end{figure}

Significant disagreement remains, however, for the larger charges,
$|\bar Q|\gesim 0.1$. In addition to the inexactness of the
comparisons for reasons already explained, another important effect
has been observed. It is that, for large charges at high
collisionality, an extended Coulomb-like (rather than Yukawa-like)
potential well appears. The $1/r^2$ potential gradient is necessary in
order to draw the ion flux to the absorbing grain through the
collisional plasma, see e.g.\cite{Su1963}. The plasma density,
potential, and velocity as a whole are then substantially perturbed by
the presence of the grain, as illustrated by Fig.\
\ref{denweb}. Notice that the density at the ($z=40$) edge of the computational
domain (Fig.\ \ref{denweb}(a)) has dropped below unity because of the
presence of the charge.
This effect is reduced by using a wider domain ($9\lambda_{De}$,
$52\times52\times128$), which indicates that it arises from the
proximity of ``mirror'' charges in the array represented by the
periodic boundary conditions.  The wider domain is used for these
high collisionality cases for $|\bar Q|\ge 0.05$, below which the effect is
negligible. Even so, the largest charges, such as shown in Fig.\
\ref{denweb}, remain substantially affected.  This is the main reason
for the upturn of the COPTIC force and the disagreement with the
analytic theory applied to the unperturbed sheath
parameters. Computational cost increases proportional to the square of
the domain width; so the calculations have not been pursued to domains
wide enough to remove perturbation at the highest charge.

\section{Application}

In order to use the dimensionless force information provided here,
either that plasma force must be translated into dimensional units or
the other forces into dimensionless form. This can be done by
realizing that the unit of force used here is
\begin{equation}
  \label{forceunit}
  F_u = n_e T_e \lambda_{De}^2= \epsilon_0 (T_e/e)^2=
  8.85\times10^{-12} T_{eV}^2,
\end{equation}
where $T_{eV}$ is the electron temperature in electron Volts. The
normalized charge (see eq.\ (\ref{grainsphere})) in dimensional units is
(to lowest order in $r_p/\lambda_{De}$)
\begin{equation}
  \label{chargedim}
  |\bar Q|= (e|\phi_p|/T_e)(r_p/\lambda_{De})  .
\end{equation}
Consequently the force normalization is 
\begin{equation}
  \label{Fuq}
  F_u|\bar Q| = \epsilon_0 T_{eV}^2 (|\phi_p|/T_{eV})(r_p/\lambda_{De}).
\end{equation}

To express a gravitational force conveniently in normalized units,
suppose that the grain density is $\rho_g$ g/cm$^3$, the acceleration
due to gravity is $g=9.8$ m/s$^2$, and let $\lambda_{mm}$ denote
$\lambda_{De}$ measured in mm, and $r_\mu$ denote $r_p$ measured in $\mu$m.
Then the force due to gravity expressed in units of $F_u|\bar Q|$ is
\begin{equation}
  \label{gravity}
  \bar{F}_g = {(4\pi/3) r_p^3\rho g\over \epsilon_0 T_{eV}^2
    (|\phi_p|/T_{eV})(r_p/\lambda_{De})} = 4.63 {\rho_g \lambda_{mm}
    r_\mu^2\over T_{eV}^2 (|\phi_p|/T_{eV})}.
\end{equation}
For example, if we consider typical dusty plasma parameters:
$\rho_g=1.5$ (melamine formaldehyde); $T_{eV}=2$; $\lambda_{mm}=0.5$;
$|\phi_p|/T_{eV}=2$; $r_\mu=3$; we get $\bar{F}_g =3.9 $. This grain
has $\bar{Q}=1.2\times10^{-2}$ and the dominant plasma force is the
background electric field. In argon the charge-exchange mean-free-path
$\ell_c$ (mm) and neutral pressure $p_n$ (Pa) are related approximately by
$\ell_cp_n\approx 6$, which determines the neutral pressure
corresponding to each collisionality. There is thus an
equilibrium between gravity and plasma force at approximately the
parameters of Table \ref{extable}.
\begin{table}[htp]
  \centering
  \begin{tabular}{|c|c|c|c|c|}
    \hline
    $\tau_c$& $p_n$ & $z/\lambda_{De}$ & $|v_f|/c_s$ & $n_i/n_{40}$ \\ \hline
    1    & 10 & 12   & 0.3 & 0.1 \\
    10   & 1  & 10   & 1.3 & 0.2 \\
    100  & 0.1& 8    & 1.9 & 0.4 \\
    \hline
  \end{tabular}
  \caption{Gravitational equilibrium values for the example case under three different collisionality levels.}
  \label{extable}
\end{table}

When the gravitational force is lower, either because the
gravitational acceleration $g$ is decreased in microgravity
environments, or because the mass density is lower for hollow spheres,
then the ion drag force becomes a more important fraction of the
equilibrium, and larger values of $|\bar{Q}|$ become important.

In summary, self-consistent one-dimensional kinetic calculations of
the DC collisional sheath have been carried out and show complicated
changes in the ion distribution function.  The plasma force on a
spherical grain of specified charge in the sheath has been found by
direct PIC simulation. It agrees quite well with the combination of
background electric field force and ion drag force. However,
differences in the drag force as much as a factor of two arise from
differences in the ion velocity distribution function. So, quantitative
agreement requires use of non-Maxwellian drift distribution in most
cases, not a shifted Maxwellian. The direct enhancement of the drag force
by collisions is also observed in strongly collisional cases. 

\subsection*{Acknowledgments}
  Useful discussions with C B Haakonsen are gratefully acknowledged.
  Work supported in part by NSF/DOE Grant DE-FG02-06ER54982.  Some of
  the computer simulations were carried out on the MIT PSFC parallel
  AMD Opteron/Infiniband cluster Loki.

\section*{Appendix: Drag force expressions}

This appendix specifies the analytic theoretical form of drag force
used to compare with the present calculations. Code to
evaluate it is available accompanying reference \cite{Hutchinson2013a}.

The drag force is written as the sum of an orbital part $F_o$ and a direct
collection part $F_c$ corrected by a collisional force factor
$\bar{F}$ equal to 1 at low collisionality; so  $F= \bar{F}(F_o+F_c)$.
The orbital part is
\begin{equation}
  \label{dragshift}
  F_o =  n_e T_e r_p^2 \left(e\phi_p\over T_e\right)^2 {T_e\over T_i} 4\pi G(u_f) \ln\Lambda,
\end{equation}
where $u_f$ is the flow velocity normalized to the ion thermal speed
$\sqrt{2T_i/m_i}$.

For the \emph{shift} distribution the function $G(u_f)$ is simply the
Chandrasekhar function
\begin{equation}
  \label{Chandra}
G_s(u)\equiv \left[{\rm erf}(u) - 2u {\rm
e}^{-u^2}/\sqrt{\pi}\right]/(2u^2)\ .  
\end{equation}
\begin{equation}
  \label{coullog}
\ln\Lambda_s=\ln\left( b_{90}+ \lambda_\ell\over  b_{90}+r_p\right),  
\end{equation}
\begin{equation}
  \label{b90s}
  b_{90s}= e\phi_p/(2T_i+{\cal E}_s) ,
\end{equation}
where
\begin{equation}
  \label{adjust}
  {\cal E}_s={0.5}m_iv_f^2\left[1+|v_f/0.4c_s|^3\right]
\end{equation}
represents the effects of flow. And
\begin{equation}\label{eq:almostlin}
\lambda_\ell^2 = r_p^2+\lambda_{De}^2/[1+T_e/(T_i+ {\cal E}_s)].
\end{equation}

For the \emph{drift} case,
\begin{equation}
  \label{Gdrift}
  G_d(u)=u/(2.66 + 1.82u^2),
\end{equation}
\begin{equation}
  \label{lnlambdad}
  \ln\Lambda_d = \ln \left( b_{90d}+ \lambda\over  b_{90d}+1.5r_p\right),
\end{equation}
 \begin{equation}
  \label{b90d}
  b_{90d}= e\phi_p/[T_i+  \sqrt{100T_iT_er_p/\lambda_{De}}\; v_f^2/(c_s^2+2.5v_f^2)],
\end{equation}
where
\begin{equation}
  \label{eq:rcutoff}
  r_c = \left(|Q|e\over 4\pi \epsilon_0 T_e r_p\right)^{1\over5}
  \left(5\sqrt{\pi}\over 8\right)^{2\over5} 
\left({r_p\over \lambda_{De}}{T_i\over T_e}\right)^{1\over5} \lambda_{De}, 
\end{equation}
\begin{equation}
  \label{eq:shiftshield}
  \lambda^{-2} =  \lambda_{De}^{-2} +(r_c^2+\lambda_{De}^2\sqrt{T_i/T_e}\;v_f/c_s)^{-1}.
\end{equation}

The collection force for the shift distribution is
\begin{equation}
  \label{Fcapprox}
  F_{cs}(u_f) = n_i T_i r_p^2 2 \pi \{u_f^2 +(1+\chi)[1-(1+bu_f)e^{-au_f}]\}
\end{equation}
where $b=0.8$,
\begin{equation}
  \label{aeq}
  a = b+ {(16+8\chi) \over 6\sqrt{\pi}(1+\chi)}
\end{equation}
and $\chi\equiv -e\phi_p/T_i$ is potential normalized to
ion temperature. For the drift distribution the collection force is
\begin{equation}
  \label{FcdFinal}
   F_{cd}(u_f) = n_i r_p^2 T_i 2 \pi \left[ 2 u_f^2 + (1+\chi) {(a-b)u_f +
     (au_f)^2 \over (1+au_f)^2}\right].
\end{equation}

The collisional correction factor is a function only of $\bar{\nu} =
r_c/\tau_cc_s$:
\begin{equation}
  \label{fcolfac}
  \bar{F}(\bar\nu) = {1 + c \bar\nu \over 1 + d \bar\nu + e \bar\nu^2}
\end{equation}
in which the coefficients are given by Table \ref{fcoltable}.
\begin{table}[htp]
\centering
\begin{tabular}{|l|c|c|c|}
\hline
  Case & $c$ & $d$ & $e$ \\
\hline
 Drift & $(7+30v_f/c_s)$ & $18v_f/c_s$ & $0.5c$ \\
 Shift & $5$ & $8v_f/c_s$ & $3.2$ \\
\hline
\end{tabular}
\caption{Coefficients for the shift and drift cases of eq.\ (\ref{fcolfac}).\label{fcoltable}
}
\end{table}

\bibliography{mybib}

\end{document}